\documentclass[iop]{emulateapj}
\usepackage{graphicx}
\usepackage{epstopdf}
\usepackage{color}

\newcommand{\Teff}{\mbox{$T_\mathrm{eff}$}}
\newcommand{\Mjup}{\mbox{$M_\mathrm{Jup}$}}
\newcommand{\Msun}{\mbox{$M_{\odot}$}}

\newcommand{\fldg}{\hbox{\sc fld-g}}

\newcommand{\vlg}{\hbox{\sc vl-g}}

\begin{document}
\shorttitle{Discovery of 2MASS~J22362452+4751425 b}
\title{Planets Around Low-Mass Stars (PALMS).  VI.  \\ Discovery of a Remarkably Red Planetary-Mass Companion to the \\ AB Dor Moving Group Candidate 2MASS~J22362452+4751425*}
\author{Brendan P. Bowler\altaffilmark{1, 2, 10},
Michael C. Liu\altaffilmark{3},
Dimitri Mawet\altaffilmark{2},
Henry Ngo\altaffilmark{2},
Lison Malo\altaffilmark{4},
Gregory N. Mace\altaffilmark{1},
Jacob N. McLane\altaffilmark{1},
Jessica R. Lu\altaffilmark{5},
Isaiah I. Tristan\altaffilmark{1, 6},
Sasha Hinkley\altaffilmark{7},
Lynne A. Hillenbrand\altaffilmark{2},
Evgenya L. Shkolnik\altaffilmark{8},
Bj\"{o}rn Benneke\altaffilmark{2},
William M. J. Best\altaffilmark{3, 9}
\\ }
\email{bpbowler@astro.as.utexas.edu}

\altaffiltext{1}{McDonald Observatory and the Department of Astronomy, The University of Texas at Austin, Austin, TX 78712, USA.}
\altaffiltext{2}{California Institute of Technology, 1200 E. California Blvd., Pasadena, CA 91125, USA.}
\altaffiltext{3}{Institute for Astronomy, University of Hawai`i at M\={a}noa; 2680 Woodlawn Drive, Honolulu, HI 96822, USA}
\altaffiltext{4}{CNRS, CFHT, 65-1238 Mamalahoa Hwy, Kamuela HI, USA}
\altaffiltext{5}{Astronomy Department, University of California, Berkeley CA 94720-3411, USA}
\altaffiltext{6}{Department of Physics and Astronomy, Rice University, MS-108, 6100 Main Street, Houston, TX 77005, USA}
\altaffiltext{7}{University of Exeter, Physics and Astronomy, EX4 4QL Exeter, UK}
\altaffiltext{8}{School of Earth and Space Exploration, Arizona State University, Tempe, AZ 85287, USA}
\altaffiltext{9}{Visiting Astronomer at the Infrared Telescope Facility, which is operated by the University of Hawaii under contract NNH14CK55B with the National Aeronautics and Space Administration}
\altaffiltext{10}{Hubble Fellow}
\altaffiltext{*}{Some of the data presented herein were obtained at the W.M. Keck Observatory, which is operated as a scientific partnership 
among the California Institute of Technology, the University of California and the National Aeronautics and Space Administration. 
The Observatory was made possible by the generous financial support of the W.M. Keck Foundation.}

\begin{abstract}

We report the discovery of an extremely red  planetary-mass companion to 
2MASS~J22362452+4751425, a $\approx$0.6~\Msun \ late-K dwarf 
likely belonging to 
the $\sim$120~Myr AB Doradus moving group.  
2M2236+4751\,b was identified in multi-epoch NIRC2 adaptive optics imaging at Keck Observatory 
at a separation of 3.7$''$, or 230 $\pm$ 20~AU in projection at the kinematic distance of 63~$\pm$~5~pc to its host star.  
Assuming membership in the AB Dor group, as suggested from its kinematics, the inferred mass of 2M2236+4751\,b is 11--14~\Mjup.
Follow-up Keck/OSIRIS $K$-band spectroscopy of
the companion reveals strong CO absorption similar to other faint red L dwarfs and lacks signs of methane absorption 
despite having an effective temperature of $\approx$900--1200~K.
With a ($J$--$K$)$_\mathrm{MKO}$ color of 2.69~$\pm$~0.12~mag, 
the near-infrared slope of 2M2236+4751\,b 
is redder than all of the HR 8799 planets and instead 
resembles
the $\approx$23~Myr isolated planetary-mass object PSO~J318.5--22,
implying that similarly thick photospheric clouds can persist in the atmospheres of giant planets
at ages beyond 100~Myr.  In near-infrared color-magnitude diagrams,
2M2236+4751\,b is located at the tip of the red L dwarf sequence 
and appears to define the ``elbow'' of the AB Dor substellar isochrone separating 
low-gravity L dwarfs from the cooler young T dwarf track. 
2M2236+4751\,b is the reddest substellar companion to a star
and will be a valuable benchmark
to study the shared atmospheric properties of young low-mass brown dwarfs and extrasolar giant planets.

\end{abstract}


\section{Introduction}{\label{sec:intro}}

L dwarfs with anomalously red near-infrared colors  were first identified 
in the Two Micron All-Sky Survey (2MASS; \citealt{Skrutskie:2006hl}) 
over a decade ago (\citealt{Dahn:2002fu})
and have steadily grown as a population ever since.
With some notable exceptions (e.g., \citealt{Looper:2008hs}; \citealt{Marocco:2014kr}), these objects were 
generally found to possess 
low-surface gravity features in their optical and near-infrared spectra,
pointing to ages much younger than the the vast majority of brown dwarfs in the field 
(\citealt{McLean:2003hx}; \citealt{Kirkpatrick:2006hb}; \citealt{Kirkpatrick:2008ec}).
In recent years the sample of red L dwarfs has ballooned 
(\citealt{Reid:2008fz}; 
\citealt{Geissler:2011gg}; \citealt{Gizis:2012kv}; 
\citealt{Liu:2013gya}; \citealt{Mace:2013jh}; \citealt{Thompson:2013kv}; 
\citealt{Schneider:2014jda}; \citealt{Marocco:2014kr}; \citealt{Best:2015em}; \citealt{Kellogg:2016fo}; \citealt{Schneider:2016iq})
as a result of all-sky infrared surveys like the Wide-field Infrared Survey Explorer (WISE),
Panoramic Survey Telescope and Rapid Response System (Pan-STARRS), 
UKIRT Infrared Deep Sky Survey (UKIDSS), and Visible and Infrared Survey Telescope for Astronomy (VISTA), 
inspiring new gravity-insensitive
spectral classification systems, all-sky searches for young brown dwarfs, and infrared parallax programs to better determine their
physical properties and relationship with young moving groups 
(\citealt{Cruz:2009gs}; \citealt{Allers:2013hk}; \citealt{Liu:2013ej}; \citealt{Gagne:2014gp}; 
\citealt{Gagne:2015ij}; \citealt{Faherty:2016fx}; \citealt{Aller:2016kg}).

It is now clear that these low-gravity brown dwarfs lie on a parallel sequence redward of
the standard L dwarf locus in near-infrared color-magnitude diagrams.
Liu et al. (2016, in press) show that the ensemble of late-M and L dwarfs 
with very low gravity spectral classifications  represent a ``fanning out'' of 
old field objects with the same spectral type;
later-type young L dwarfs are progressively redder and fainter in $M_J$ than their higher-gravity counterparts.
This systematic offset and broadening of the L dwarf locus 
is likely caused by a combination of youthful overluminosity for late-M and early-L dwarfs and 
unusually dusty photospheres at later types 
that redistribute emergent flux to longer wavelengths (e.g., \citealt{Filippazzo:2015dv}).
While this narrative may be qualitatively correct, no set of self-consistent atmospheric and evolutionary 
models can completely reproduce the colors and spectra of this fascinating population.

In parallel with discoveries of young isolated brown dwarfs, adaptive optics imaging surveys 
have identified a similar pattern of redder colors and fainter absolute 
magnitudes among young extrasolar giant planets (see, e.g., \citealt{Bowler:2016jk}).
The archetypal companion in this class is 2M1207--3932~b (\citealt{Chauvin:2004cy});
with a $J$--$K$ color of 3.1~$\pm$0.2~mag (\citealt{Dupuy:2012bp}), it is the reddest ultracool object known and 
lacks the deep methane absorption expected at the effective temperature predicted by
cooling models (\citealt{Mohanty:2007er}; \citealt{Patience:2010hf}).  
More generally, this empirically-discovered trait--- not predicted beforehand by atmospheric models---
is also seen in other red, low-temperature companions like  
HR~8799 bcde (e.g., \citealt{Marois:2008gm}; \citealt{Bowler:2010ft}; 
\citealt{Barman:2011fe}; \citealt{Skemer:2014hy}; \citealt{Bonnefoy:2016gx}), 
2M0122--2439~B (\citealt{Bowler:2013ek}; \citealt{Hinkley:2015gk}), 
and VHS~J1256--1257~b (\citealt{Gauza:2015fw}),
as well as young free-floating brown dwarfs near the canonical L/T transition (e.g., \citealt{Liu:2013gya}) and may be caused
by vigorous vertical mixing and strong 
disequilibrium carbon chemistry (\citealt{Barman:2011dq}; \citealt{Zahnle:2014hl}).

A natural question that has emerged from these studies is the degree to which young isolated 
brown dwarfs and extrasolar giant planets share common atmospheric properties
and cooling pathways.  The physical properties of the lowest-mass brown 
dwarfs appear to broadly overlap with the most massive giant planets.
However, if the initial conditions or compositions of these populations are substantially different
from one another as a result of disparate formation channels then this similarity may be superficial
and free-floating young red L dwarfs may not be ``exoplanet analogs''  after all.
Unfortunately, the dearth of known red L dwarf companions currently  prevents
this kind of comparison.

Here we present the discovery of an extraordinarily red companion to the late-K dwarf
2MASS~J22362452+4751425.
The host star was identified by \citet{Schlieder:2012gj} as a candidate member
of the AB Dor moving group on the basis of its proper motion and UV emission from $GALEX$.
Our new radial velocity measurements of the primary are consistent with AB Dor membership,
implying a mass of $\sim$11--14~\Mjup \ for the companion.
At a separation of 3$\farcs$7 (230~AU), 2MASS~J22362452+4751425~b
(herinafter 2M2236+4751\,b) belongs to two classes of rare and enigmatic objects: 
planetary-mass companions on extremely wide orbits ($>$100~AU), which have low occurrence rates
of $<$2\% around young stars (\citealt{Bowler:2016jk}), and 
red L dwarf companions, of which only a handful are known.
2M2236+4751\,b is likely to be an important benchmark for both populations.

\section{Observations}{\label{sec:observations}}

\begin{deluxetable*}{lccccccc}
\tabletypesize{\scriptsize}
\setlength{ \tabcolsep } {.12cm} 
\tablewidth{0pt}
\tablecolumns{8}
\tablecaption{Keck/NIRC2 Adaptive Optics Imaging of 2MASS~J22362452+4751425 \label{tab:obs}}
\tablehead{
 \colhead{UT Date} &  \colhead{Epoch}  & \colhead{$N$$\times$Coadds$\times$$t_\mathrm{exp}$}  & \colhead{Filt.}                 & \colhead{Sep.}      & \colhead{P.A.}              & \colhead{$\Delta$mag}  & \colhead{FWHM\tablenotemark{a}}  \\
 \colhead{}          &  \colhead{(UT)}       & \colhead{(s)}                  & \colhead{}                      &  \colhead{(mas)}       & \colhead{ ($^{\circ}$)} &  \colhead{}    & \colhead{(mas)}                
        }
\startdata
2014 Nov 08  & 2014.852  &  5 $\times$ 1 $\times$ 60  &  $K_S$+cor600  &  3692 $\pm$ 3  &  135.8 $\pm$ 0.2  &  7.3 $\pm$ 0.5 & 69 $\pm$ 9  \\
2015 Aug 27  & 2015.653  &  5 $\times$ 1 $\times$ 60  &  $K_S$+cor600  &  3694 $\pm$ 3  &  135.7 $\pm$ 0.2  &  8.22 $\pm$ 0.04 & 48  $\pm$ 1  \\
2016 Jun 27  & 2016.489  &  10 $\times$ 1 $\times$ 60  &  $K_S$+cor600  &  3696 $\pm$ 3  &  135.3 $\pm$ 0.2  &  8.2 $\pm$ 0.3 &   $\cdots$   \\
2016 Jun 27  & 2016.489  &  10 $\times$ 1 $\times$ 60  &  $H$+cor600  &  3705 $\pm$ 3  &  134.8 $\pm$ 0.2  &  9.8 $\pm$ 0.4 &  41 $\pm$ 1  \\
2016 Jul 18  & 2016.546  &  10 $\times$ 10 $\times$ 6  &  $H$+cor600  &  3707 $\pm$ 3  &  135.4 $\pm$ 0.2  &  9.0 $\pm$ 0.2  & $\cdots$  \\
2016 Jul 19  & 2016.549  &  10 $\times$ 6 $\times$ 10  &  $J$+cor600  &  3705 $\pm$ 3  &  135.7 $\pm$ 0.2  &  10.4 $\pm$ 0.3 &  34 $\pm$ 1 \\
2016 Aug 03  & 2016.590  &  2 $\times$ 1 $\times$ 60  &  $J$+cor600  &  3690 $\pm$ 3  &  135.4 $\pm$ 0.2  &  9.93 $\pm$ 0.12   &  36 $\pm$ 1
\enddata
\tablenotetext{a}{FWHM of 2M2236+4751\,A as measured from unsaturated frames.}
\end{deluxetable*}


\begin{figure}
  \vskip -.6 in
  \hskip -1.2 in
  \resizebox{6.5in}{!}{\includegraphics{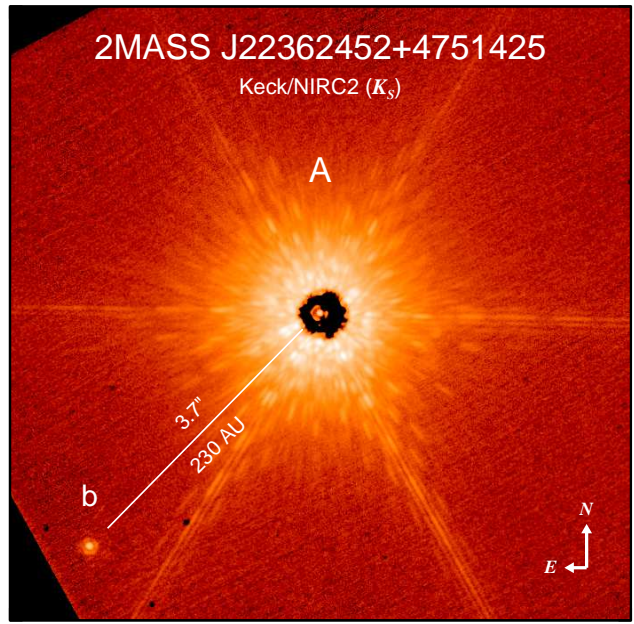}}
  \vskip -.6in
  \caption{Single NIRC2 $K_S$-band image of  2M2236+4751\,Ab from August 2015.  
  The K7 host star is positioned behind the 
  partly translucent 600-mas diameter coronagraph.  2M2236+4751\,b is located at a separation of 3$\farcs$7 
  near the edge of the array in this 7$''$$\times$7$''$ image.  
  North is up and East is to the left.  \label{fig:nirc2img} } 
\end{figure}


\begin{figure*}
  \vskip -1.9 in
  \hskip -0.75 in
  \resizebox{8in}{!}{\includegraphics{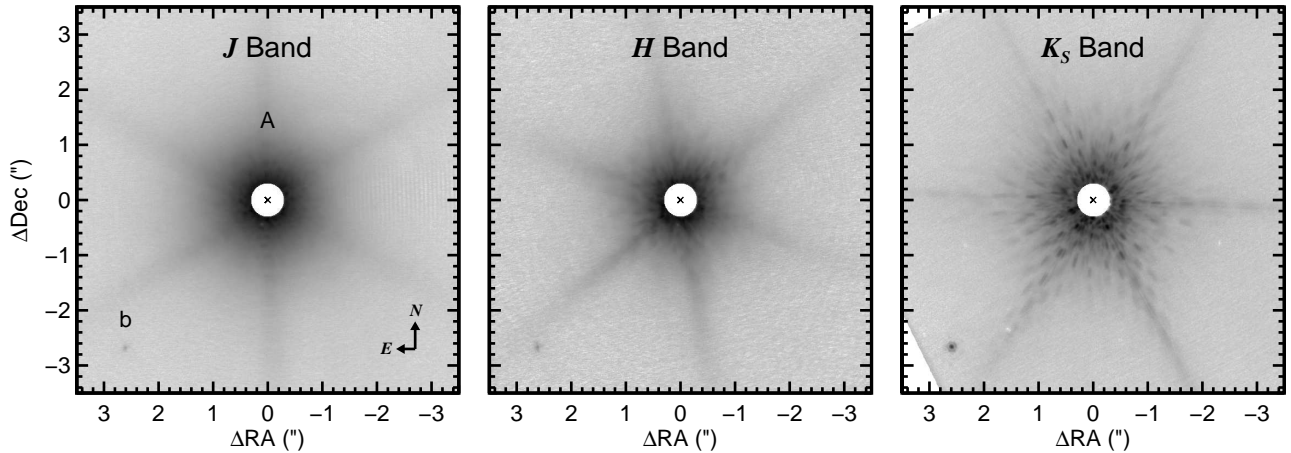}}
  \vskip -1.4 in
  \caption{Median-combined NIRC2 images of 2M2236+4751\,Ab in $J$, $H$, and $K_S$ filters 
  from July 2016, June 2016, and August 2015, respectively.  
  2M2236+4751\,b is unusually red with a $J$--$K_S$ color of 2.7~mag.
  Individual frames were derotated before being combined and north aligned to correct for a small amount of rotation
  experienced in pupil-tracking mode.  This resulted in a slight azimuthal blurring of the host star, which is 
  especially evident in $J$ and $H$ bands.  The position of the host star is marked with an ``x.'' 
  Images have been stretched with an inverse hyperbolic sine transformation.  North is up and East is to the left.  \label{fig:jhkimgs} } 
\end{figure*}

\subsection{Keck/NIRC2 Adaptive Optics Imaging}{\label{sec:obs:igrins}}

We first imaged 2M2236+4751 on 2014 November 8 UT in $K_S$-band with Keck/NIRC2 using natural guide star adaptive optics
(NGS AO; \citealt{Wizinowich:2013dz}) as part of our ongoing efforts to find, characterize, and measure the statistical properties 
of giant planets around young low-mass stars.
2M2236+4751\,b was identified at a separation of 3$\farcs$7 (Figure~\ref{fig:nirc2img}) and subsequently confirmed to be comoving with its host
star through follow-up imaging in 2015 and 2016 (Table~\ref{tab:obs}).  All observations were carried out in a
similar manner with the narrow camera using the entire 1024$\times$1024 pix$^2$ array, the partly-translucent 600~mas diameter coronagraph, 
and vertical angle (pupil-tracking) rotator setup.
In addition to coronagraphic imaging, we also acquired 10--20 unsaturated frames of the host star 
for photometric calibration immediately before or 
after the deeper imaging 
except on 2016 June 27 UT ($K_S$ filter) and 2016 July 18 UT ($H$ filter); for these data sets we use the 
coronagraph transmission measurement
from \citet{Bowler:2015ja} to derive the flux ratios of 2M2236+4751\,Ab.

Raw images were bias-subtracted, flat-fielded, and corrected for bad pixels and cosmic rays.  
Field rotation was small (between 2--6$^{\circ}$) in our 5--10~min sequences and the companion is outside the speckle noise-limited
region close to the star, so no angular differential imaging subtraction was necessary.  
Each frame was corrected for optical distortions using the solution from \citet{Yelda:2010ig} for data obtained prior to April 2015 and
from \citet{Service:2016gk} thereafter to account for the altered optical distortion following a NIRC2 pupil realignment.
The corresponding plate scales and north orientation angles are 9.952 $\pm$ 0.002 mas pix$^{-1}$ and +0.252 $\pm$ 0.009$^{\circ}$
from Yelda et al., and 9.971 $\pm$ 0.004 mas pix$^{-1}$ and +0.262 $\pm$ 0.02$^{\circ}$ from Service et al.
Images were registered using the position of the host star visible behind the occulting mask and were then de-rotated, median-combined, 
and north-aligned to produce a final reduced frame (Figure~\ref{fig:jhkimgs}).

Astrometry of the companion is derived in a similar manner as in \citet{Bowler:2015ja}
and takes into account positional uncertainties of the star behind the coronagraph,
centroid errors of the companion, and systematic errors in the distortion solution and 
north alignment, the latter of which is the dominant term in the error budget.
When unsaturated frames were obtained, relative photometry was measured by calculating the mean and standard deviation 
of counts from the host star in the unsaturated images and from the companion in the deep 
coronagraphic data using aperture photometry.
Photometric uncertainties are propagated analytically.  Our NIRC2 astrometry and relative photometry is listed in Table~\ref{tab:obs}.

\subsection{Keck/OSIRIS Spectroscopy of 2M2236+4751\,b}{\label{sec:obs:osiris}}

We obtained a medium-resolution ($R$ $\equiv$ $\lambda$/$\Delta \lambda$ $\approx$ 3800) 1.96--2.38~$\mu$m 
spectrum of 2M2236+4751\,b with the OH-Suppressing Infra-Red Imaging Spectrograph 
(OSIRIS; \citealt{Larkin:2006jd}) mounted on Keck I and coupled with NGS AO on 2016 June 23 UT.
Our observations benefited from a new grating installed in 2012 (\citealt{Mieda:2014dt}) and 
a new spectrograph detector in early 2016, a few months prior to our observing run.
Observations were taken with the $Kbb$ filter and the 50~mas pix$^{-1}$ plate scale, 
resulting in a 16$\times$64 spaxel$^2$ (spatial pixel) lenslet geometry 
and 0$\farcs$8$\times$3$\farcs$2 rectangular field of view.
The rotator was set orthogonal to the binary position angle (P.A.) to avoid contamination from the host star.
We acquired six pairs of nodded data cubes in an AB pattern with individual exposures of 300~s, totaling 60~min of
on-source integration time altogether.
Conditions were clear with 0$\farcs$4 seeing throughout our science observations.
Immediately prior to this we observed the A0V standard HD172728 for telluric correction.

The raw 2D images were transformed to 3D data cubes with the OSIRIS data reduction pipeline and the 
latest rectification matrices from Keck.  Spectra were extracted using aperture photometry with 
local sky subtraction then scaled to a common level and median-combined.  
Telluric correction was carried out with the \texttt{xtellcor\_general} routine in the 
Spextool reduction package for IRTF/SpeX (\citealt{Vacca:2003wi}; \citealt{Cushing:2004bq}).
Finally, our spectrum was flux calibrated using a multiplicative scale factor corresponding to the 
apparent $K_S$ magnitude measured with NIRC2.

\begin{deluxetable*}{lcccccccc}
\tabletypesize{\scriptsize}
\setlength{ \tabcolsep } {.1cm} 
\tablewidth{0pt}
\tablecolumns{9}
\tablecaption{Spectroscopic Observations\label{tab:specobs}}
\tablehead{
        \colhead{Object}   &  \colhead{Date}      &   \colhead{Telescope/}  &  \colhead{Filter}  &  \colhead{Slit Width}  & \colhead{Plate Scale}   &    \colhead{Tot. Exp.}   &   \colhead{Resolution} & \colhead{Standard\tablenotemark{a}}   \\
                     \colhead{}  &  \colhead{(UT)}   &   \colhead{Instrument}       &  \colhead{}           &  \colhead{($''$)}               & \colhead{(mas pix$^{-1}$)}             &    \colhead{(min)}   & \colhead{(=$\Delta \lambda$/$\lambda$)}  &   \colhead{}             
        }   
\startdata
2M2236+4751\,A    &    2016 Jun 13     &   CFHT/ESPaDOnS  &  $\cdots$  &   $\cdots$  &  $\cdots$   &   15  & 68000  &  $\cdots$  \\
2M2236+4751\,A    &    2016 Jun 14     &   McDonald 2.7-m/IGRINS  &  $\cdots$  &   0.98  &  $\cdots$   &   32  & 45000  &  HD 219290  \\
2M2236+4751\,A    &    2016 May 24    &   IRTF/SpeX (SXD)  &  $\cdots$  &    0.3 & $\cdots$  &   4  & 2000  &  HD 209932  \\
2M2236+4751\,b    &    2016 Jun 23     &   Keck/OSIRIS  &  $Kbb$  &    $\cdots$ & 50  &   60  & 3800  &  HD 172728  
\enddata
\tablenotetext{a}{Radial velocity standard or telluric standard.}
\end{deluxetable*}

\subsection{CFHT/ESPaDOnS Optical Spectroscopy}{\label{sec:obs:espadons}}

High-resolution optical spectroscopy was obtained for 2MJ2236+4751\,A in queue service observing 
(QSO) mode with ESPaDOnS (\citealt{Donati:2006vj}) at CFHT
on 2016 June 13 UT. ESPaDOnS was used in the Star+Sky spectroscopic mode and combined with the normal CCD readout mode to yield a resolving power of R$\sim$68000 covering the
3700 to 10500 \AA \ wavelength range. The total integration time was 900 seconds. 
The data were reduced by the QSO team using the CFHT pipeline  
UPENA1.0, which uses the Libre-ESpRIT software package (\citealt{Donati:1997wj}).
A heliocentric radial velocity of --22.1 $\pm$ 0.5 km s$^{-1}$ and a projected rotational velocity of 4 km s$^{-1}$ were measured using the same methodology as described in \citet{Malo:2014dk}.

\subsection{Harlan J. Smith Telescope/IGRINS Spectroscopy}{\label{sec:obs:igrins}}

We observed 2M2236+4751\,A with the Immersion Grating Infrared
Spectrometer (IGRINS) on the 2.7-meter Harlan J. Smith telescope
located at McDonald Observatory. IGRINS is a high-resolution
near-infrared spectrograph offering
simultaneous coverage of the $H$ and $K$ bands (1.45-2.45 $\mu$m) at
a resolution of R$\simeq$45,000 (\citealt{Park:2014kn}). Observations consisted of 
four 480 s exposures taken in an ABBA nod
pattern. Immediately after we observed the A0V star HD 219290
at a similar airmass for telluric correction.

The IGRINS data were reduced using version 2.1 of the IGRINS reduction
pipeline\footnote{\url{https://github.com/igrins/plp}}
\citep{lee16}. The pipeline performs dark, bias, and flat
corrections, followed by an optimal extraction of the
one-dimensional spectrum for both the target and A0V star. A first
order wavelength solution is generated using lines from a ThAr lamp;
a full wavelength solution is then generated using OH emission sky lines. Finally,
a more refined wavelength solution is calculated using
telluric absorption lines in the raw A0 spectrum. The final spectrum has a S/N
of $\sim$150 per resolution element. 

The broad spectral grasp of IGRINS provides more than 20,000 resolution elements at R$\simeq$45,000. 
Cross-correlation of all of these elements provides robust statistical measurements of stellar radial velocities.
As described in \citet{Mace:2016ev}, the spectral stability of IGRINS is sub-pixel within an observing night, producing 
a radial velocity precision of $<$ 200 m/s. 
Radial velocities have been derived for nearly all IGRINS observations. 
Briefly, the IGRINS spectrum of 2M2236+4751\,A was cross-correlated in pixel space against 144 other IGRINS spectra with spectral types between K3 and M2. 
Shifts in the spectrum are removed by cross-correlating the telluric spectra to determine pixel shifts between the target and the templates with similar spectral type.
Pixel offsets between stellar spectra are converted into velocities using the empirically derived spectral resolution. 
The relative radial velocity is determined from the mean and standard error 
of all radial velocity measurements from 
all 144 template spectra, after a sigma filter that removes $>$4$\sigma$ outliers.
The relative radial velocity is converted into an absolute radial velocity using a zero-point shift based on $>$100 absolute radial velocities from the literature.
Our final barycentric radial velocity for 2M2236+4751\,A is --21.4 $\pm$ 0.2 km s$^{-1}$,
consistent with (and more precise than) the CFHT/ESPaDOnS measurement. \\

\subsection{IRTF/SpeX Near-Infrared Spectroscopy}{\label{sec:obs:igrins}}

To better characterize the host star we obtained a medium-resolution 0.7--2.55~$\mu$m  
spectrum of 2M2236+4751\,A on 2016 May 24 UT with the short cross-dispersed mode of the recently-upgraded 
SpeX (\citealt{Rayner:2003vo}) spectrograph 
at the NASA Infrared Telescope Facility (IRTF).
The 0$\farcs$3 slit was used which resulted in an average resolving power of $R$$\approx$2000.
Four pairs of 30-s exposures were taken in an ABBA pattern along the 15$''$ slit.  
The A0V standard HD~209932 was observed immediately beforehand at a similar airmass.
Images were pairwise subtracted then the spectra were extracted, 
median-combined, and corrected for telluric features
using Spextool reduction package (\citealt{Vacca:2003wi}; \citealt{Cushing:2004bq}).

\section{Results}{\label{sec:observations}}


\begin{figure}
  \vskip -.25 in
  \hskip -.25 in
  \resizebox{3.8in}{!}{\includegraphics{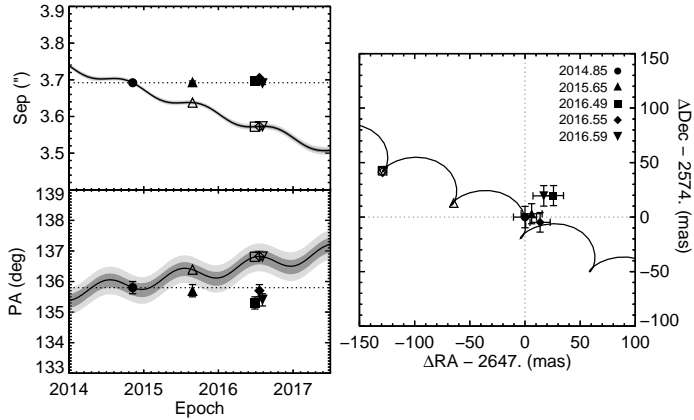}}
  \vskip -.1 in
  \caption{Expected relative motion of a stationary background object (solid curve) based on the proper and 
  parallactic motion of 2M2236+4751\,A.  Our measured astrometry (filled symbols) is consistent with a constant
  separation and position angle (dotted lines) for 2M2236+4751\,b over time, indicating it is comoving with its host star.  
  Open symbols show the expected positions of 2M2236+4751\,b if it were stationary for each epoch of our observations.  Gray shaded
  regions denote 1-$\sigma$ and 2-$\sigma$ confidence intervals for the background tracks which incorporate uncertainties
  in our first epoch astrometry, kinematic distance, and host star proper motion.      \label{fig:backtracks} } 
\end{figure}


\begin{figure*}
  \vskip -.4 in
  \hskip .1 in
  \resizebox{7in}{!}{\includegraphics{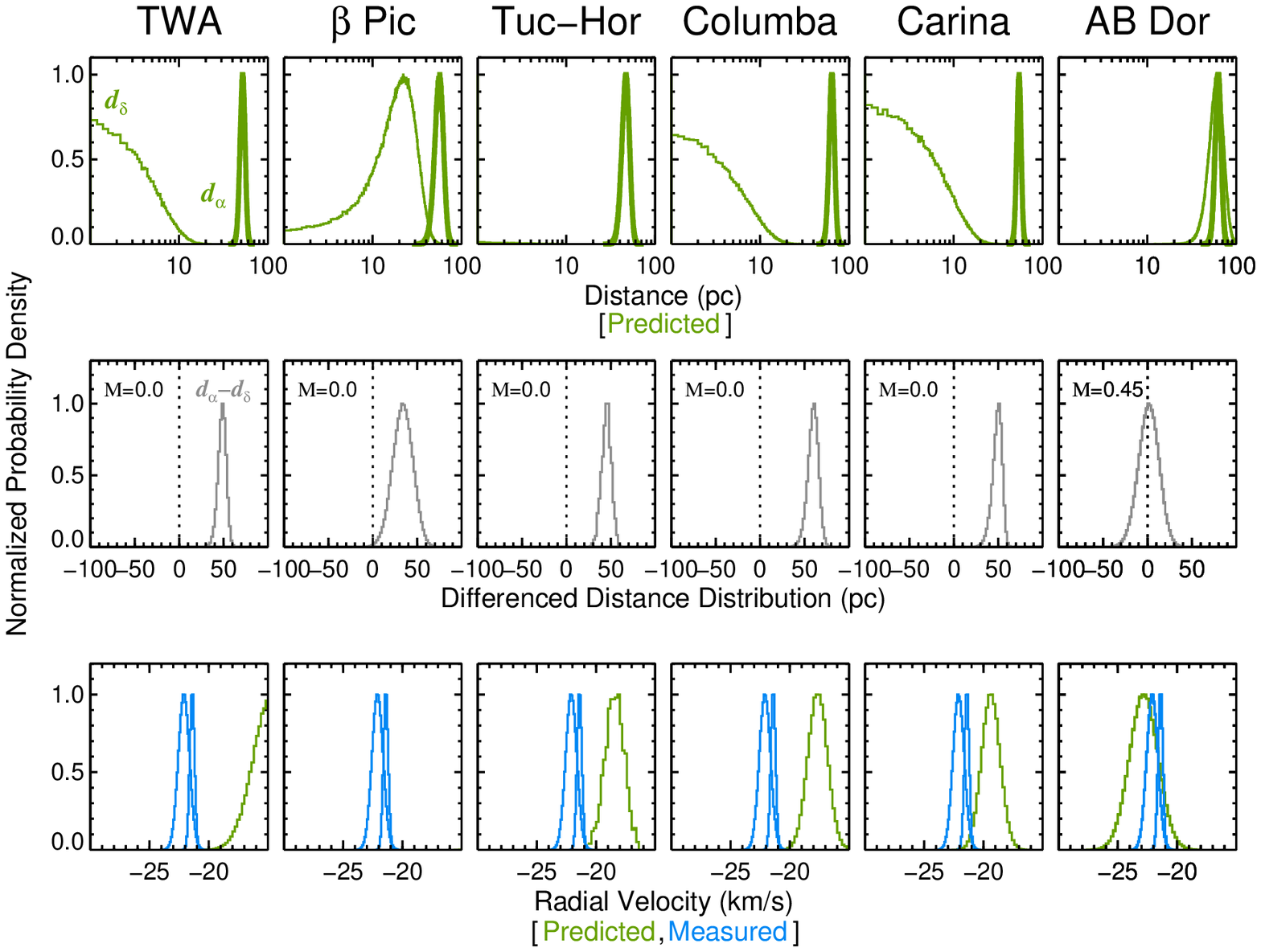}}
  \vskip -.4 in
  \caption{Young moving group membership tests based on the proper motion of 2M2236+4751.  By adopting moving group $UVW$ 
  space velocities from \citet{Torres:2008vq}, the measured proper motion and sky position of 
  2M2236+4751 can be inverted to solve for the expected distance (for both right ascension 
  and declination proper motion components; top row) and the expected radial velocity (bottom row)
  assuming group membership.  The middle row shows the difference of the two distance distributions;
  the agreement of the two are represented by the metric $\mathcal{M}$, which is the lower tail integral of the 
  differenced distribution about zero (Bowler et al., in prep.).  2M2236+4751 agrees well with AB Dor but is inconsistent with 
  all other moving groups tested here.  Moreover, the measured radial velocity from ESPaDOnS and IGRINS (blue, bottom row)
  are consistent with the predicted value.   \label{fig:d3plot} } 
\end{figure*}

\subsection{Common Proper Motion}{\label{sec:cpm}}

Our NIRC2 astrometry of 2M2236+4751\,Ab spans 1.7 years (Table~\ref{tab:obs}).  
Throughout this period, the 
expected change in separation and P.A. for a stationary background star are 119 $\pm$ 5 mas and 1.0 $\pm$ 0.3$^{\circ}$, 
respectively,
based on our first epoch of astrometry, the host star's proper motion from UCAC4 
($\mu_{\alpha}$cos$\delta$ = 62.6 $\pm$ 1.4 mas yr$^{-1}$, $\mu_{\delta}$=--30.5 $\pm$ 1.9 mas yr$^{-1}$; \citealt{Zacharias:2013cf}),
and its kinematic distance (63~$\pm$~5 pc; Section \ref{sec:age}).
The measured change between the initial and latest epochs is only 2~$\pm$ 4~mas and 0.4 $\pm$ 0.3$^{\circ}$, 
heavily favoring the common proper motion scenario (Figure~\ref{fig:backtracks}).

The comoving  versus stationary background hypotheses can be tested more
quantitatively using the Bayes factor.  Following \citet{Bowler:2013ek}, the $\chi^2$ values
of the comoving ($M_1$) and stationary ($M_2$) scenarios are 30 and 2850, respectively, for five degrees of freedom.
Assuming equivalent prior odds, the posterior odds are $\log$($P(M_1)$/$P(M_2)$) = 613, 
indicating the pair are unambiguously comoving and very likely to be gravitationally bound.

\subsection{Age and Young Moving Group Membership}{\label{sec:age}}

The age of 2M2236+4751 is critical to the interpretation of this system; a young age would imply
that the companion has a mass in the planetary regime and its red colors are a 
result of thick clouds associated with low surface gravity like 2M1207--3932~b,
while an old age would indicate that it is a brown dwarf with 
an unusually dusty (and potentially metal-rich) photosphere like 2M2148+4003 (\citealt{Looper:2008hs}).
Here we discuss two lines of evidence regarding the system age: a potential kinematic link
to the $\approx$120~Myr AB Dor moving group, and other age indicators like activity and rotation period.

\citet{Schlieder:2012gj} identified 2M2236+4751 as a candidate member of the AB Dor moving group
from its proper motion and sky position.  We also find consistency with AB Dor using an alternative method to identify
new members of young moving groups (Bowler et al., in prep.).  
By adopting the $UVW$ kinematics and uncertainties of moving groups 
from \citet{Torres:2008vq}, we can invert the standard calculation of galactic space velocities (\citealt{Johnson:1987ji})
when only a proper motion is available to predict an expected radial velocity distribution as well as two distance distributions 
associated with each proper motion 
component ($\mu_{\alpha}$cos$\delta$ and $\mu_{\delta}$).  These two distance distributions
should agree with one another if the object is a member of a particular group being tested; subtracting one from the
other and calculating the tail integral about zero offers a simple metric ($\mathcal{M}$) to assess this consistency.
$\mathcal{M}$ values near 0.5 are most consistent, while incompatible distributions have $\mathcal{M}$ values near 0.  

For 2M2236+4751 we find an $\mathcal{M}$ value of 0.45 for AB Dor and values of 0.0 for the 
TWA, $\beta$~Pic, Tuc-Hor, Columba, and Carina moving groups, indicating excellent proper
motion consistency with the AB Dor moving group (Figure~\ref{fig:d3plot}).  
To assess the false positive probability that non-moving group members
would have $\mathcal{M}$ values at least this high by chance, we ran the same analysis for 
a sample of over 2000 inactive M dwarfs from \citet{Gaidos:2014if}.  The probability that a field star would share
a similarly consistent proper motion with the AB Dor moving group is 1.1\%.  We adopt this as an upper limit because 
the joint probability of a similar proper motion \emph{and} radial velocity (see below) will be substantially smaller than this.
Note that one advantage of our method over Bayesian techniques is that our approach 
does not rely on probabilistic assignments using kinematic ($UVW$) and 
positional ($XYZ$) models for moving groups and field stars.  
For example, the BANYAN~I algorithm developed by \citet{Malo:2013gn} 
predicts a probability of 68\% for AB Dor membership, but BANYAN~II
(which uses updated membership lists and is tailored to low-mass moving group members; \citealt{Gagne:2014gp}) 
gives an AB Dor probability of only 0.11\%.\footnote{The BANYAN I and II AB Dor membership probabilities
increase to 84\% and 0.9\%, respectively, when also using the weighted mean radial velocity.  The low probability 
from BANYAN II is likely based 
on the $Y$ position of 2M2236+4751 compared to bona fide AB Dor members, 
which we discuss in more detail later in this section.  For reference, note that the 
BANYAN II ``young field'' and ``old field'' probabilities are 52\% and 47\%.}

We find predicted (kinematic-based) radial velocity and distances
of --22.8~$\pm$~1.2 km s$^{-1}$ and 63 $\pm$ 5 pc.
This is in good agreement with our measured radial velocities from ESPaDOnS (--22.1 $\pm$ 0.5 km s$^{-1}$) and
IGRINS (--21.4 $\pm$ 0.2 km s$^{-1}$), bolstering a possible kinematic association with AB Dor.
Similarly, the photometric distance to 2M2236+4751 is 74~$\pm$~10 pc 
based on the $M_V$ versus $V$--$K_S$ relationship
for the Pleiades from \citet{Bowler:2013ek}, which is $\approx$1$\sigma$ from the kinematic distance.
Our results are also consistent with the radial velocity and distance predictions 
of --23.0 $\pm$ 1.2 km s$^{-1}$ and 65 $\pm$ 9 pc inferred by Schlieder et al.

\begin{figure*}
  \vskip -.4 in
  \hskip .4 in
  \resizebox{6in}{!}{\includegraphics{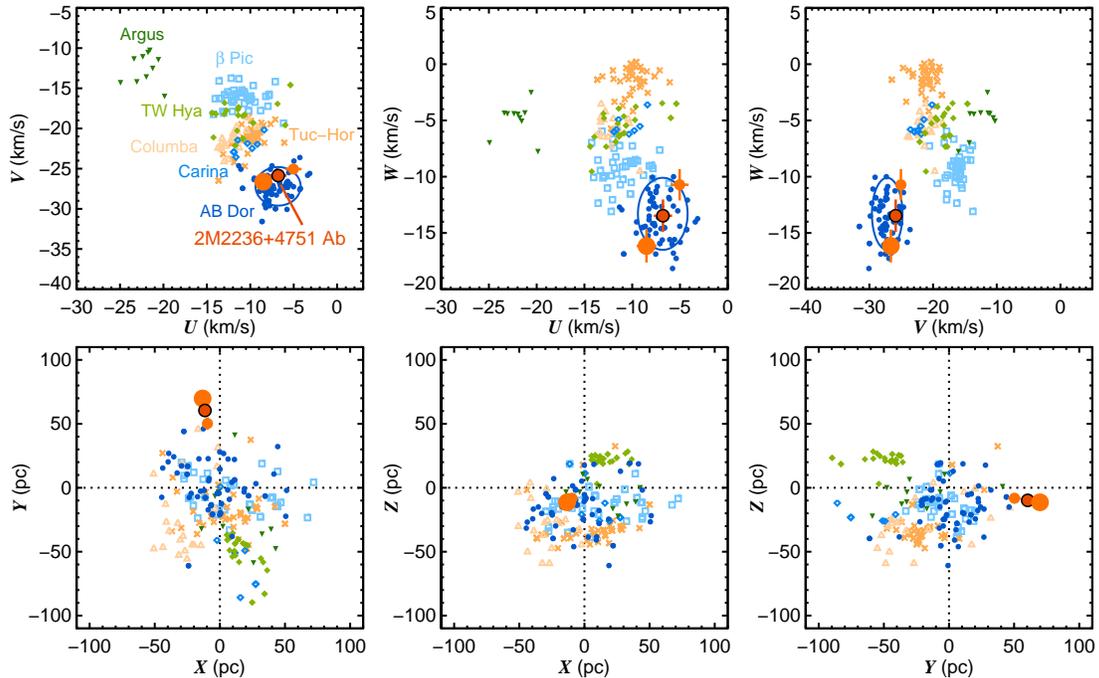}}
  \vskip -.4 in
  \caption{Partially-constrained $UVW$ galactic space velocities (top) and $XYZ$ heliocentric positions (bottom) for 2M2236+4751\,Ab
  compared to nearby young moving groups.  Red circles denote distances of 53 $\pm$ 5 pc, 63 $\pm$ 5, and 73 $\pm$ 5 pc to 2M2236+4751\,A
  based on its proper motion and weighted mean radial velocity.  The space velocities of 2M2236+4751\,Ab are in good agreement with 
  known AB Dor members from \citet{Gagne:2014gp} and the AB Dor locus from \citet[2-$\sigma$ ellipses]{Torres:2008vq}.
  2M2236+4751\,Ab is similar to known members in $X$ and $Z$ and lies just beyond the established population in the $Y$ direction.
   \label{fig:uvw} } 
\end{figure*}

Figure~\ref{fig:uvw} shows the $UVW$ galactic space velocities and $XYZ$ heliocentric positions for 
2M2236+4751 based on its proper motion, weighted mean radial velocity (--21.5 $\pm$ 0.2 km s$^{-1}$),
and kinematic distance compared to confirmed members of nearby young moving groups from \citet{Gagne:2014gp}.
The agreement with known AB Dor members in $U$, $V$, $W$, $X$, and $Z$ is excellent, but 
the $Y$ position appears to be an outlier.  It is unclear how problematic this is since the current census
of moving group members is incomplete and heavily biased towards nearby ($\lesssim$60~pc) 
young stars owing to
distance cuts in early searches (see, e.g., \citealt{Zuckerman:2004ku}).  Soon $Gaia$ will
reveal the entire stellar population and $UVW$/$XYZ$ distributions of these nearby moving groups.
Until then, we simply note this possible 
tension with the $Y$ positional distribution of AB Dor members and await a re-evaluation in the future.

Activity diagnostics offer additional independent ways to constrain the age of this system.  
2M2236+4751 was not detected in the \emph{Roentgen SATellite} ($ROSAT$) All-Sky Survey (RASS).  
Adopting an upper limit equal to the detection limit of
RASS (0.05 cnts s$^{-1}$; \citealt{Voges:1999ws}) and assuming a Hardness
Ratio 1 (HR1) value of 0.0, the implied X-ray flux from 2M2236+4751 is $f_X$ $<$ 4.1$\times$10$^{-13}$ 
erg cm$^{-2}$ s$^{-1}$, the X-ray luminosity is log($L_X$) $<$ 29.3 erg s$^{-1}$, and the fractional 
X-ray to bolometric luminosity is log($L_X$/$L_\mathrm{bol})$ $<$ --3.1 dex.\footnote{If instead we
adopt an HR1 value of --0.3, which is the typical value for older field stars (\citealt{Bowler:2012cs}),
the resulting upper limits are $f_X$ $<$ 3.4$\times$10$^{-13}$ 
erg cm$^{-2}$ s$^{-1}$, log($L_X$) $<$ 29.2 erg s$^{-1}$, 
and log($L_X$/$L_\mathrm{bol})$ $<$ --3.2 dex.}
These values are near the saturation limit for X-ray emission in low-mass stars,
so a non-detection in RASS is not particularly useful to constrain the coronal
emission and age of 2M2236+4751.
Low-mass Pleiades members span a wide range of X-ray luminosities,   
log($L_X$/$L_\mathrm{bol})$ $\approx$ --3 to --4 dex (\citealt{Preibisch:2005jz}),
so this X-ray upper limit is consistent with a Pleiades (and AB-Dor)-like age of $\sim$120~Myr.
Table~\ref{tab:k7} lists the activity levels for known or suspected K7 members of AB~Dor;
like Pleiades members, these objects have fractional X-ray luminosities between --3.0 and --4.0 dex,
consistent with our upper limit for 2M2236+4751.

\begin{deluxetable*}{lcccccccccc}
\tabletypesize{\scriptsize}
\setlength{ \tabcolsep } {.1cm} 
\tablewidth{0pt}
\tablecolumns{11}
\tablecaption{K7 Dwarfs in the AB Dor Moving Group \label{tab:k7}}
\tablehead{
        \colhead{Object}   &  \colhead{$\alpha_\mathrm{2000.0}$}      &   \colhead{$\delta_\mathrm{2000.0}$}  & \colhead{$d$} &  \colhead{log($L_X$/$L_\mathrm{bol}$)\tablenotemark{a}}  &  \colhead{$EW$(H$\alpha$)}  & \colhead{$EW$(Li)}   &    \colhead{$P_\mathrm{rot}$}  &  \colhead{$NUV$--$J$} & \colhead{$FUV$--$J$}  & \colhead{Ref}  \\
 & \colhead{(h m s)}  &  \colhead{($^{\circ}$ $'$ $''$)} & \colhead{(pc)}  &  \colhead{(dex)}  & \colhead{(\AA)} &  \colhead{(m\AA)} & \colhead{(days)} & \colhead{(mag)}  & \colhead{(mag)}
        }   
\startdata  
2M0034+2523 & 00 34 08.43 & +25 23 49.8 & 48                  &  --3.23 $\pm$ 0.10  &  --0.82       &  $\cdots$     &    3.16  &   10.12 $\pm$ 0.07  &  12.3 $\pm$ 0.4           &   1, 2, 3 \\
HIP 25283  &  05 24 30.17   & --38 58 10.7  &  18  &  --3.64 $\pm$ 0.04  &  $\cdots$ & 12  &   9.34 & $\cdots$                &  $\cdots$                  &   4, 5, 6 \\
HIP 26369  &  05 36 55.10  &  --47 57 48.1  &  26       &  --3.36 $\pm$ 0.21  &  --0.9        &  70  &  4.54                       &   $\cdots$                 &    $\cdots$              &   4, 5, 7 \\
HIP 31878 & 06 39 50.04 & --61 28 41.8  &  22.4  &  --3.82 $\pm$ 0.15  & $\cdots$  & 50 &  9.06  &   10.41 $\pm$ 0.03 & 13.1 $\pm$ 0.13       &   4, 6, 7, 8  \\
HIP 86346  &  17 38 39.65 & +61 14 16.1  &  33        & --3.04 $\pm$ 0.06   &  --1.3         & 40 & 1.842 &  9.36 $\pm$ 0.04  & 11.14 $\pm$ 0.03  &   4, 8, 9, 10 \\
2M2039+0620  &  20 39 54.60  &  +06 20 11.8  &  38.5              &  --3.95 $\pm$ 0.12    &    0.3        &  $\cdots$  &  $\cdots$  &   $\cdots$  &  $\cdots$                                &     1, 2 \\
HIP 106231  &  21 31 01.71 & +23 20 07.5  &  24.8  & --3.10 $\pm$ 0.09  &  $\cdots$ & 215                          &   0.423                      &   9.04 $\pm$ 0.03    &  $\cdots$              &  4, 7, 11 \\
HIP 113597     &  23 00 27.92  &  --26 18 43.17  &  30  &  --3.68       &    --0.1   & 10   &   8.0  &  10.47 $\pm$ 0.02  &  12.97 $\pm$ 0.13          &   4 ,  11, 12 , 13 \\
\hline
2M2236+4751       &  22 36 24.52   &  +47 51 42.5    &   63  &  $<$3.1            &   0.18   &  $<$50   &    11.2      &  10.9 $\pm$ 0.2    &   $\cdots$                            &    14, 15
\enddata
\tablenotetext{a}{Fractional X-ray luminosities are computed using $ROSAT$ count rates following \citet{Bowler:2013ek}.}
\tablerefs{
(1) \citet{McCarthy:2012hk};
(2) \citet{Lepine:2013hc};
(3) \citet{Norton:2007im};
(4) \citet{vanLeeuwen:2007dc};
(5) \citet{Torres:2006bw};
(6) \citet{Messina:2010kx};
(7) \citet{Kiraga:2012wj};
(8) \citet{Fernandez:2008hu};
(9) \citet{Gizis:2002ej};
(10) \citet{Henry:1995ip};
(11) \citet{daSilva:2009eu}; 
(12) \citet{Riaz:2006du};
(13) \citet{Messina:2011kz};
(14) This work;
(15) \citet{Hartman:2011cq}.
}
\end{deluxetable*}


\begin{figure}
  \vskip -.1 in
  \hskip -.4 in
  \resizebox{4in}{!}{\includegraphics{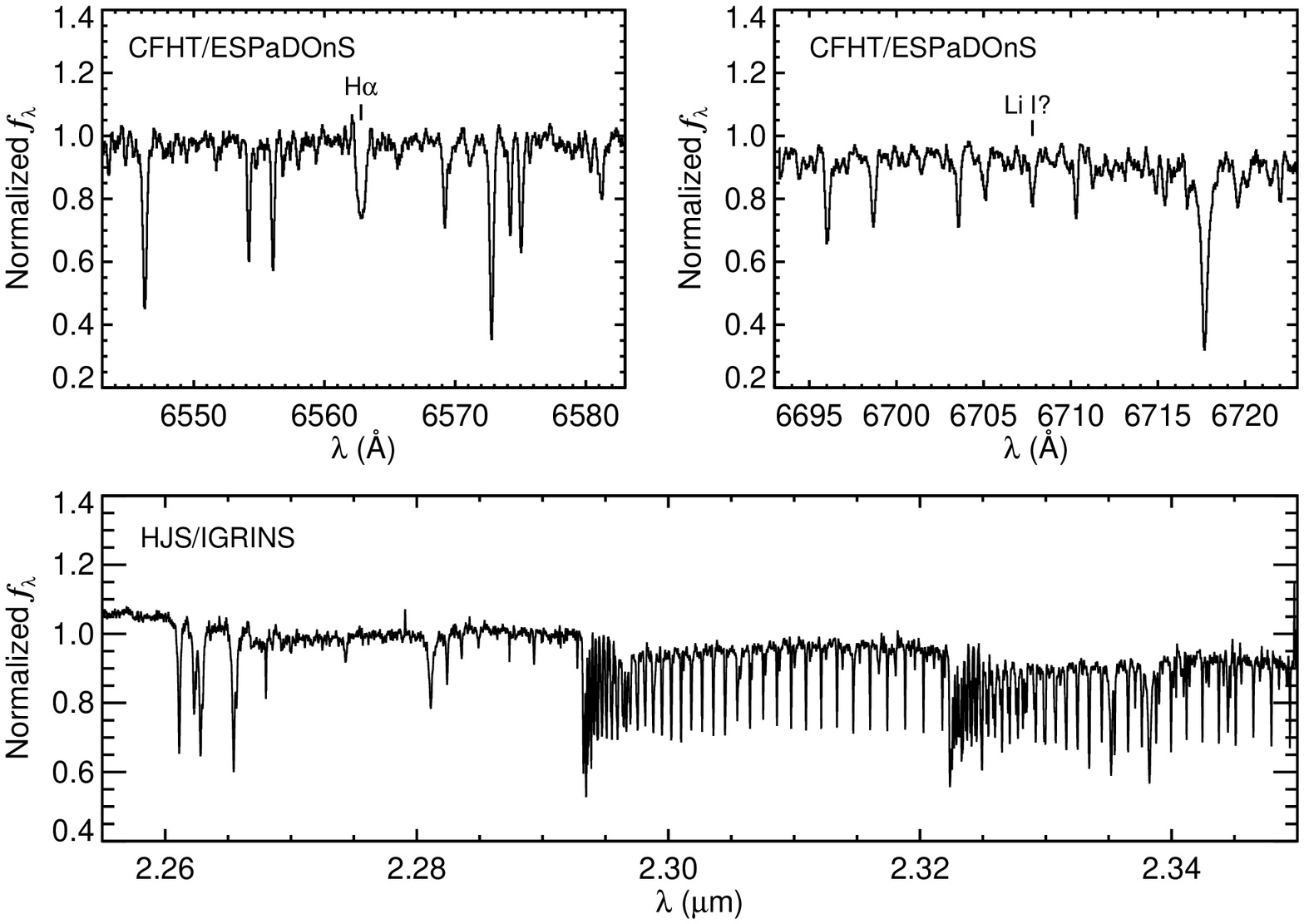}}
  \vskip -.2 in
  \caption{High-resolution optical and near-infrared spectra from CFHT/ESPaDOnS (top panels) and
  HJS/IGRINS (bottom), respectively.  2M2236+4751\,A shows H$\alpha$ absorption and possible  
  \ion{Li}{1} doublet absorption at 6707.8~\AA, 
  although this may instead originate from \ion{Fe}{1} at 6707.4~\AA.
  The $^{12}$CO $\nu$=2--0 and $\nu$=3--1 band heads at 2.2935 $\mu$m and 2.3227 $\mu$m are visible in our IGRINS data.
  The ESPaDOnS spectrum has been shifted by +22.1~km s$^{-1}$ to account for its radial velocity.\label{fig:hostspec} } 
\end{figure}

UV photometry from the \emph{Galaxy Evolution Explorer}
($GALEX$; \citealt{Martin:2005ko}; \citealt{Morrissey:2007ch}) 
provides another age diagnostic for  2M2236+4751.
The host star is detected in the near-UV filter ($NUV$ = 20.9 $\pm$ 0.2 mag) but not in the far-UV ($FUV$).
\citet{Findeisen:2011ha} show that older stars trace an envelope in $NUV$--$J$ color
as a function of near-infrared color; the $NUV$--$J$ color of 10.9 $\pm$ 0.2 mag for 2M2236+4751
is generally consistent with Hyades-like chromospheric emission but slightly redder
than known K7 members in AB Dor with $GALEX$ detections (Table~\ref{tab:k7}).
Similarly, the F$_{FUV}$/F$_{J}$ ratio (9 $\pm$ 2 $\times$10$^{-5}$) from \citet{Shkolnik:2011in}
and (F$_{FUV}$/F$_{J}$)$_\mathrm{exc}$ excess ratio above the photosphere (8 $\pm$ 2 $\times$10$^{-5}$) 
from \citet{Shkolnik:2014jl} indicate that 2M2236+4751 has lower chromospheric emission levels than most star at AB Dor-like
ages.  However, the large intrinsic scatter makes it difficult to rule out the all but the
youngest ages ($\lesssim$10~Myr) for this object.

The H$\alpha$ equivalent width measured from our high-resolution ESPaDOnS spectrum 
is +0.18 $\pm$ 0.02 \AA \ in absorption (Figure~\ref{fig:hostspec}).  A lack of H$\alpha$ emission is consistent with 2M2236+4751's
apparently weak chromospheric and coronal emission.  This value is also somewhat weaker than the
envelope traced out by older main sequence stars ($\approx$--0.6~\AA \ for a $B$--$V$ color of 1.2~mag; see, e.g., 
Figure~5 of \citealt{Zuckerman:2004ku}), indicating the H$\alpha$ line may be slightly filled in for this star.
H$\alpha$ emission for K7 AB Dor members rangers from +0.3 to --1.3 (Table~\ref{tab:k7}).
A weak absorption line at about 6707.8 \AA \ may be caused by either the \ion{Li}{1} doublet or 
 possibly \ion{Fe}{1} at 6707.4~\AA.
 If it originates from lithium, its equivalent width  ($EW$=40 $\pm$ 10 m\AA)
is comparable to other AB Dor members with similar optical colors (e.g., \citealt{Torres:2008vq}).  

Stars spin up as they contract to the zero-age main sequence and then slowly spin down over time through
magnetic braking and winds, so in 
principle rotation periods offer another tool to constrain stellar ages.
In practice, age determinations using rotation periods are imprecise for low-mass stars 
because they exhibit a large spread in initial angular momenta that continues to broaden 
over time (e.g, \citealt{Irwin:2011gz}; \citealt{Mcquillan:2014gp}).  
\citet{Hartman:2011cq} measure a period of 11.2~days for 2M2236+4751; compared to the 
Pleaides rotation distribution, which has a comparable age to AB Dor and a well-characterized
stellar population, this period sits near the maximum envelope of rotation rates 
for stellar masses of $\approx$0.6~\Msun \ 
(\citealt{Hartman:2010dv}; \citealt{Covey:2016ix}; \citealt{Rebull:2016wb}).
K7 members in AB Dor exhibit a similarly broad range of rotation periods from
0.4--9.3 days (Table~\ref{tab:k7}).


\begin{figure}
  \vskip -1.3 in
  \hskip -1 in
  \resizebox{5.5in}{!}{\includegraphics{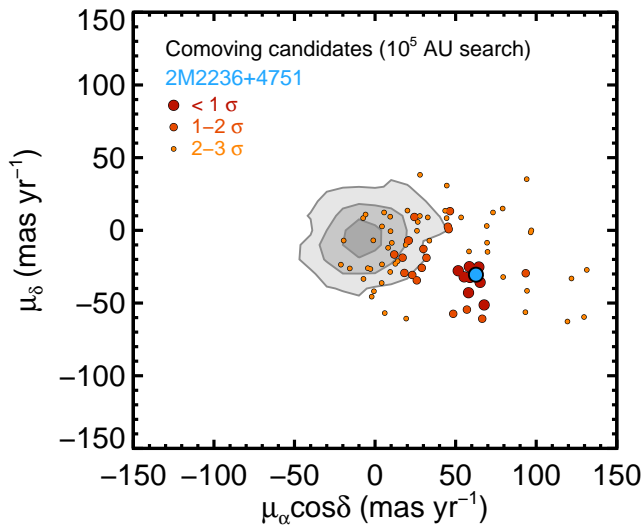}}
  \vskip -3.3 in
  \caption{Search for wide stellar companions in the PPMXL proper motion catalog out to 10$^5$~AU 
  from 2M2236+4751.  Gray shaded regions centered at ($\mu_{\alpha}$$\cos \delta$, $\mu_{\delta}$) = (0, 0) mas yr$^{-1}$
  show contours encapsulating 68\%, 90\%, and 95\% of the field population in the direction of 2M2236+4751.
  Candidates within 1-, 2-, and 3-$\sigma$ of 2M2236+4751 are shown as red and orange filled circles.
  Although 78 stars are consistent within 3-$\sigma$, none of those with $V$-band magnitudes fall on or above the main
  sequence at the distance of 2M2236+4751.  
   \label{fig:pmpm} } 
\end{figure}


\begin{figure}
  \vskip -.3 in
  \hskip -1.4 in
  \resizebox{7in}{!}{\includegraphics{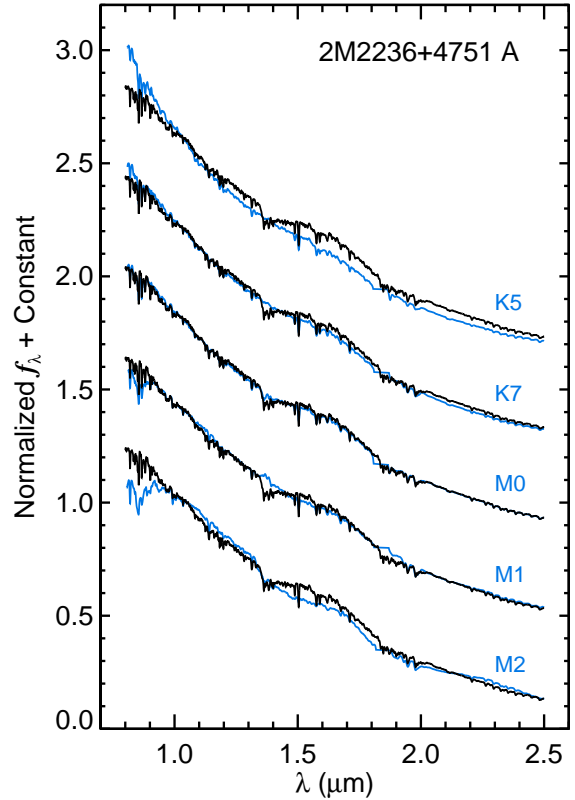}}
  \vskip -.3 in
  \caption{IRTF/SpeX near-infrared spectrum of 2M2236+4751\,A (black) compared with K5--M2 dwarf templates from
  the IRTF Spectral Library (blue; \citealt{Rayner:2009ki}).  2M2236+4751\,A shows slight steam absorption at $\sim$1.4 and
  $\sim$1.9 $\mu$m and is most similar to the M0 template.   However, the optical colors of the host are slightly bluer
  than expected for M0 so we adopt a [K7 $\pm$ 1] classification.   \label{fig:hostcomp} } 
\end{figure}

Altogether we find that the kinematics and rotation period of 2M2236+4751 are 
consistent with the AB Dor moving group.  There is some tension between the heliocentric galactic 
$Y$ position compared to accepted members, and similarly the activity level of 2M2236+4751 
appears to be lower than AB Dor members of the same spectral type, 
but neither of these decisively rule out membership.
We conclude that 2M2236+4751 is either an inactive member of AB Dor, presumably 
in the quiet tail end of the activity distribution, or it is an older kinematic interloper
in this group.  The unusually red spectrum of the companion may suggest the system
is indeed a member of this group, similar to the dusty L dwarf WISE~J0047+6803 (\citealt{Gizis:2012kv}; \citealt{Gizis:2015ey}).
Because of the excellent kinematic agreement with AB Dor, 
we adopt the cluster age of $\approx$120~Myr for this work 
(\citealt{Luhman:2005kx}; \citealt{Barenfeld:2013bf})
but note that older ages are also possible if further investigation 
shows it not a member.

\subsection{Search for Wide Stellar Companions}{\label{sec:cpm}}

Wide stellar companions to exoplanet host stars provide a more 
complete view of the system architecture
and offer an independent way to age-date the planet (e.g., \citealt{Mamajek:2012ga}).
We searched for companions to 2M2236+4751\,Ab by querying the PPMXL
proper motion catalog (\citealt{Roeser:2010cr}) at separations corresponding to 5000 AU, 10$^4$ AU, 
and 10$^5$ AU.  \citet{Wu:2011jm} found that PPMXL proper motions 
have systematic errors of $\approx$--2 $\pm$ 5 mas yr$^{-1}$ based on a large calibration using 
stationary quasars.  We therefore add this systematic uncertainty
in quadrature with all objects returned in our search, then compute the difference in proper motions
between 2M2236+4751 and stars within each specified angular radius (Figure~\ref{fig:pmpm}).  
Uncertainties are propagated analytically and all stars within 3-$\sigma$ are 
considered to be candidate comoving companions.

Real stellar companions should sit on or above the main sequence at the distance to 
2M2236+4751, so we also impose a color-magnitude diagram cut to further vet the sample.
$V$ and $K_S$-band magnitudes are queried from UCAC4;
objects falling below the main sequence as determined by \citet{Pecaut:2013ej}\footnote{Absolute $V$-band magnitudes compiled by E. Mamajek can be found at http://www.pas.rochester.edu/ $\sim$emamajek/EEM\_dwarf\_UBVIJHK\_colors\_Teff.txt.}
in $M_V$ versus $V$--$K_S$  are  excluded from consideration.

Aside from the host star itself, no candidate companions emerged in our 5000~AU and 10$^4$~AU searches.
Within 10$^5$ AU, 77 stars have proper motions within $<$3$\sigma$ of 2M2236+4751. 
Based on the same analysis for an offset field 2$^{\circ}$ away, 
the expectation value for the number of stars having
consistent proper motions by chance is 34 and the probability of 
finding at least one star consistent with 2M2236+4751 is effectively 1.0 from binomial statistics.
However, none of these candidates also have $V$-band magnitudes and fall on or above the main 
sequence.  The probability of at last one star passing all these criteria by chance (absolute $V$-band magnitude 
above the main sequence and consistent proper motion)
based on our offset field is only 0.28.
Altogether, no promising stellar companions emerged in our search.

\subsection{Physical Properties of 2M2236+4751\,A}{\label{sec:cpm}}

No spectral type for 2M2236+4751\,A is evident in the literature, 
but its optical through near-infrared colors suggest $\sim$K7.  For example, 
its $B$--$V$ color of 1.2~$\pm$~0.2~mag, $V$--$K_S$ color of 3.36 $\pm$ 0.03~mag, 
$g'$--$r'$ color of 0.59 $\pm$ 0.09 mag are all consistent with K7--M0 types 
(\citealt{Tokunaga:2000tr}; \citealt{Drilling:2000vo}; \citealt{Bochanski:2007it}).
This agrees with the estimate of K7 from \citet{Lepine:2011gl} based on $V$--$J$ color.
Furthermore, our SpeX spectrum of 2M2236+4751\,A shows slight steam absorption at
$\sim$1.4~$\mu$m and $\sim$1.9~$\mu$m and is most similar to M0 template from
the IRTF Spectral Library (Figure~\ref{fig:hostcomp}; \citealt{Rayner:2009ki}).
Altogether we estimate a spectral type of [K7 $\pm$ 1], where the brackets indicate it is 
predominantly photometrically-based.

We estimate a bolometric luminosity of $\log(L_\mathrm{bol}$/$L_{\odot})$ = --1.17 $\pm$ 0.08 dex for
the host star using the $H$-band bolometric
correction from \citet{Casagrande:2008dt} and its kinematic distance.  Similarly, we estimate an
effective temperature of 4045 $\pm$ 35 K from their \Teff($V$--$J$) relation.
The corresponding mass using the \citet{Baraffe:2015fwa} evolutionary models and 
based on the bolometric luminosity and age (120~Myr) is 0.60 $\pm$ 0.05 \Msun.
This inferred mass is independent of older ages so the same value is expected if this system is a 
kinematic interloper in AB Dor.


\begin{figure}
  \vskip -.4 in
  \hskip -.7 in
  \resizebox{5in}{!}{\includegraphics{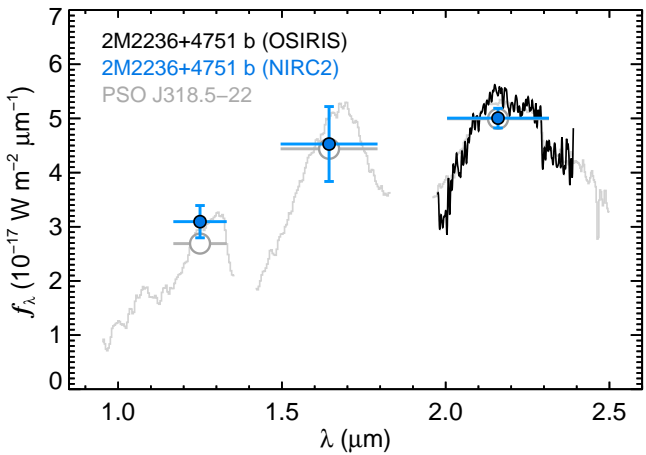}}
  \vskip -.6 in
  \caption{Near-infrared spectral energy distribution of 2M2236+4751\,b.  Blue points display our NIRC2 photometry of the companion,
  and our OSIRIS spectrum (smoothed from $R$$\sim$3800 to $R$$\sim$1000) is shown in black.  
  The near-IR spectrum of PSO~J318.5--22 from \citet{Liu:2013gya} is shown in gray for comparison 
  along with synthetic (scaled) photometry.     \label{fig:vhscomp} } 
\end{figure}


\begin{figure*}
  \vskip -.4 in
  \hskip 0.3 in
  \resizebox{7in}{!}{\includegraphics{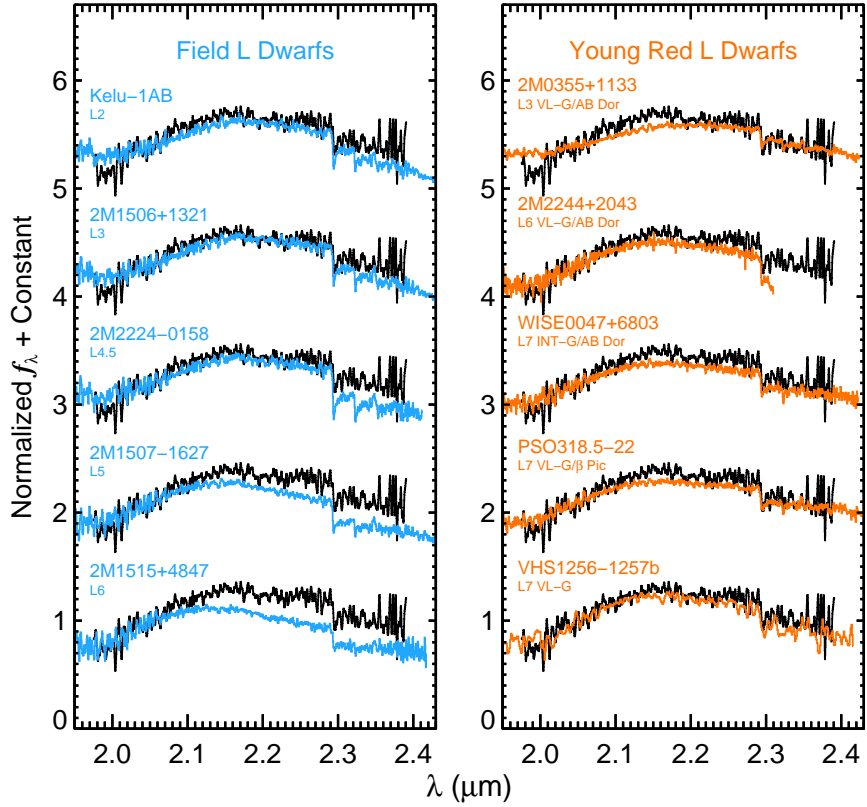}}
  \vskip -.4 in
  \caption{$K$-band spectrum of 2M2236+4751\,b compared with field L dwarfs (blue; left panel) and young red L
  dwarfs (orange; right panel).  The overall 2.0--2.1~$\mu$m slope and CO strength is most consistent with 
  red late-L dwarfs like PSO~J318.5--22 and VHS~J1256--1257~b, though broader coverage is needed for a 
  more accurate classification.  Comparison spectra are from \citet{Cushing:2005ed} and \citet{Rayner:2009ki} 
  for field L dwarfs and \citet{Allers:2013hk}, \citet{McLean:2003hx}, \citet{Gizis:2015ey}, \citet{Liu:2013gya},
  and \citet{Gauza:2015fw} for red L dwarfs.       \label{fig:speccomp} } 
\end{figure*}

\subsection{Photospheric and Physical Properties of 2M2236+4751\,b}{\label{sec:cpm}}

With a $J$--$K_S$ color of 2.62 $\pm$ 0.12 mag and a $J$--$H$ color of 1.4 $\pm$ 0.2 mag, 
the near-infrared spectral energy distribution of 2M2236+4751\,b is among the 
reddest L dwarfs known.  The $J$--$K_S$ color in particular rivals the most extreme objects
in the color-magnitude diagram like 2M1207--3932~b (\citealt{Chauvin:2004cy}), 
PSO~J318.5--22 (\citealt{Liu:2013gya}), HR~8799~b (\citealt{Marois:2008ei}),
 and VHS~J1256--1257~b (\citealt{Gauza:2015fw})
whose unusual spectra are thought to be caused by extremely thick photospheric clouds.

Figure~\ref{fig:vhscomp} shows our NIRC2 photometry and our $K$-band OSIRIS spectrum of 
2M2236+4751\,b.  The 1.2--2.2~$\mu$m spectral shape and $K$-band spectrum closely
match the near-infrared spectrum of PSO~J318.5--22, a 5--8~\Mjup \ L7 \vlg \ member of 
$\beta$~Pic (\citealt{Liu:2013gya}; \citealt{Allers:2016gz}).
The 2.3~$\mu$m $^{12}$CO bandhead and red 2.0--2.2~$\mu$m 
spectral slope are evident in the OSIRIS data.  
Compared to field L dwarfs and young red L dwarfs in Figure~\ref{fig:speccomp},
the overall $K$-band shape of 2M2236+4751\,b more closely resembles 
the latter population of objects with very low and intermediate gravity classifications.
\citet{Allers:2013hk} show that a $K$-band spectrum alone is not
sufficient for precise spectral and gravity classifications of young L dwarfs;
lacking broader spectral coverage, we adopt ``late-L pec" for 2M2236+4751\,b.

\begin{deluxetable}{lccc}
\tabletypesize{\scriptsize}
\setlength{ \tabcolsep } {.12cm} 
\tablewidth{0pt}
\tablecolumns{4}
\tablecaption{Properties of the 2MASS~J22362452+4751425 Ab System}
\tablehead{
   \colhead{Property} & \colhead{2M2236+4751\,A}  &  \colhead{2M2236+4751\,b}    & \colhead{Ref} 
        }       
\startdata
\cutinhead{Astrometry and Kinematics}
$\mu_{\alpha}$cos$\delta$ (mas yr$^{-1}$) & 62.6 $\pm$ 1.4  & $\cdots$  &  1   \\
$\mu_{\delta}$  (mas yr$^{-1}$)  & --30.5 $\pm$ 1.9 & $\cdots$ &  1 \\
$d_\mathrm{kin}$ (pc)\tablenotemark{a} & 63 $\pm$ 5 & $\cdots$  &  2  \\
$d_\mathrm{phot}$ (pc) & 74 $\pm$ 10 & $\cdots$  &  2  \\
$RV_\mathrm{kin}$ (km s$^{-1}$)\tablenotemark{a} & --22.8 $\pm$ 1.2 & $\cdots$  & 2  \\
$RV_\mathrm{ESPaDOnS}$ (km s$^{-1}$) &  --22.1 $\pm$ 0.5 &  $\cdots$ & 2 \\
$RV_\mathrm{IGRINS}$ (km s$^{-1}$) & --21.4 $\pm$ 0.2  &  $\cdots$ & 2 \\
$\overline{RV}$ (km s$^{-1}$)\tablenotemark{b} &  --21.5 $\pm$ 0.2 &  $\cdots$ & 2 \\
$UVW$ (km s$^{-1})$\tablenotemark{c}  &  \multicolumn{2}{c}{--6.9 $\pm$ 1.0, --25.9 $\pm$ 0.4, --13.7 $\pm$ 1.5}  &  2  \\
$XYZ$ (pc)\tablenotemark{c}  & \multicolumn{2}{c}{--11.6 $\pm$ 0.9, 61.1 $\pm$ 4.8, --10.0 $\pm$ 0.8}  &  2 \\

\cutinhead{Photometry}
$\Delta J$ (mag)\tablenotemark{d} & \multicolumn{2}{c}{9.99 $\pm$ 0.11} & 2  \\
$\Delta H$ (mag)\tablenotemark{d} & \multicolumn{2}{c}{9.16 $\pm$ 0.18} & 2  \\
$\Delta Ks$ (mag)\tablenotemark{d} & \multicolumn{2}{c}{8.2 $\pm$ 0.04} & 2 \\
$GALEX$ $NUV$  (mag) &  20.9 $\pm$ 0.2  &  $\cdots$  &  3   \\
$B$ (mag) &  13.73 $\pm$ 0.18 & $\cdots$ &  4 \\
$V$ (mag) &  12.51 $\pm$ 0.03 & $\cdots$ &  4 \\
$g'_\mathrm{APASS}$ (mag) &  13.1 $\pm$ 0.2 & $\cdots$ &  4  \\
$r'_\mathrm{APASS}$ (mag) &  11.91 $\pm$ 0.03 & $\cdots$ &  4  \\
$i'_\mathrm{APASS}$ (mag) &  11.32 $\pm$ 0.08 & $\cdots$ &  4  \\
$R2$ (mag) &  11.5 & $\cdots$ &  5 \\
$J_\mathrm{MKO}$ (mag) & 9.975 $\pm$ 0.022\tablenotemark{e} &  19.97 $\pm$ 0.11  &  2, 6  \\
$H_\mathrm{MKO}$ (mag) & 9.388 $\pm$ 0.021\tablenotemark{e}  &  18.54 $\pm$ 0.18  &  2, 6  \\
$K_\mathrm{MKO}$ (mag) & [9.180 $\pm$ 0.018]\tablenotemark{f}  & [17.28 $\pm$ 0.04]\tablenotemark{f}  &  2, 6   \\
$K_s$ (mag) & 9.148 $\pm$ 0.018  & 17.35 $\pm$ 0.04  &  2, 6   \\
$W1$  (mag) &  9.058 $\pm$ 0.023  & $\cdots$  & 7   \\
$W2$  (mag) & 9.094  $\pm$ 0.020  & $\cdots$  & 7 \\
$W3$  (mag) &  8.947 $\pm$ 0.027  & $\cdots$  & 7 \\
$W4$  (mag)  & 8.731 $\pm$ 0.356 &  $\cdots$  & 7   \\
$(J-H)_\mathrm{MKO}$ (mag) & 0.59 $\pm$ 0.03 &  1.4 $\pm$ 0.2  &  2, 6  \\
$(H-K)_\mathrm{MKO}$ (mag) & [0.132 $\pm$ 0.001] &  [1.26 $\pm$ 0.18]  &  2, 6  \\
$(J-K)_\mathrm{MKO}$ (mag) & [0.747 $\pm$ 0.001] &  [2.69 $\pm$ 0.12]  &  2, 6  \\
$H_\mathrm{MKO}-Ks$ (mag) & 0.24 $\pm$ 0.03 &  1.19 $\pm$ 0.18  &  2, 6  \\
$J_\mathrm{MKO}-Ks$ (mag) & 0.827 $\pm$ 0.03 &  2.62 $\pm$ 0.12  &  2, 6  \\
$M_{J_\mathrm{MKO}}$ (mag)\tablenotemark{g}  & 5.97 $\pm$ 0.17 &  15.97 $\pm$ 0.2 & 2 \\
$M_{H_\mathrm{MKO}}$ (mag)\tablenotemark{g} & 5.39 $\pm$ 0.17 &  14.5 $\pm$ 0.3 & 2 \\
$M_{K_S}$ (mag)\tablenotemark{g} & 5.15 $\pm$ 0.17 &  13.35 $\pm$ 0.18  &  2  \\
$M_{K_\mathrm{MKO}}$ (mag)\tablenotemark{g} & [5.18 $\pm$ 0.17] &  [13.28 $\pm$ 0.18]  &  2  \\

\cutinhead{Physical Properties}
Separation  ($''$) & \multicolumn{2}{c}{3.70}   &  2 \\
Separation  (AU)\tablenotemark{g} & \multicolumn{2}{c}{230 $\pm$ 20}  &  2\\
Age (Myr)\tablenotemark{a} &   \multicolumn{2}{c}{120 $\pm$ 10}   &  8  \\
log($L_\mathrm{bol}$/$L_{\odot}$)  &  --1.17 $\pm$ 0.08 &  --4.57 $\pm$ 0.06  &  2 \\
Mass   &  0.60 $\pm$ 0.05 \Msun  &  11--14~\Mjup   &  2  \\
Spectral Type &   [K7 $\pm$ 1]\tablenotemark{h}   &  late-Lpec  &  2   \\
\enddata
\tablenotetext{a}{Assumes membership in AB Dor.}
\tablenotetext{b}{Weighted mean of IGRINS and ESPaDOnS radial velocities.}
\tablenotetext{c}{$U$ and $X$ are positive toward the galactic center, $V$ and $Y$ are positive toward the
direction of galactic rotation, and $W$ and $Z$ are positive toward the north galactic pole.}
\tablenotetext{d}{Weighted mean of relative photometry from Table~\ref{tab:obs}.}
\tablenotetext{e}{Assumes $J_\mathrm{MKO}$$\approx$$J_\mathrm{2MASS}$ and $H_\mathrm{MKO}$$\approx$$H_\mathrm{2MASS}$ for 2M2236+4751\,A.}
\tablenotetext{f}{Based on synthetic $K_S$--$K_\mathrm{MKO}$ color from our SpeX spectrum of the primary
($K_S$--$K_\mathrm{MKO}$ = --0.0320 $\pm$ 0.0008 mag) and our OSIRIS spectrum of the companion 
($K_S$--$K_\mathrm{MKO}$ = 0.070 $\pm$ 0.004 mag).}
\tablenotetext{g}{Assumes kinematic distance to the host.}
\tablenotetext{h}{Estimated spectral type based on photometry.}
\tablerefs{
(1) UCAC4 (\citealt{Zacharias:2013cf});
(2) This work;
(3) \citet{Morrissey:2007ch};
(4) APASS (\citealt{Henden:2016uu});
(5) USNO-B1.0 (\citealt{Monet:2003bw});
(6) 2MASS (\citealt{Cutri:2003tp});
(7) WISE (\citealt{Cutri:2012tx});
(8) \citet{Torres:2008vq}.
}
\end{deluxetable}

A shallow indentation is visible in the pseudo-continuum of our OSIRIS spectrum from $\approx$2.2--2.3~$\mu$m 
that is similar to the onset of methane absorption in the latest field L dwarfs.
To more quantitatively assess whether this feature is indeed from methane, we compared our spectrum
with molecular templates of pure H$_2$O, CO, and CH$_4$ following
\citet{Konopacky:2013jvc} and \citet{Barman:2015dy}.
The templates were generated by computing a customized thermal emission model 
using the SCARLET atmospheric retrieval framework  (\citealt{Benneke:2015tw}). 
The model self-consistently calculates the thermal structure and equilibrium chemistry 
at $T_\mathrm{eff}$ = 1200~K for a cloud-free atmosphere with solar elemental composition 
at a resolving power of $R$ $>$ 250,000.\footnote{The SCARLET model considers the molecular opacities of 
H$_2$O, CH$_4$, NH$_3$, HCN, CO, CO$_2$ and TiO from the high-temperature 
ExoMol database (\citealt{Tennyson:2012ca});
O$_2$, O$_3$, OH, C$_2$H$_2$, C$_2$H$_4$, C$_2$H$_6$, H$_2$O$_2$, and HO$_2$ 
from the HITRAN database (\citealt{Rothman:2009ea}); 
and H$_2$-broadening following the prescription in \citet{Burrows:2003ty}. 
Collision- induced broadening from H$_2$/H$_2$ and H$_2$/He collisions 
is computed following \citet{Borysow:2002iv}.}
The molecular templates are then convolved with the instrument profile, 
flattened by fitting a high-order polynomial, and cross-correlated with the 
our flattened $K$-band spectrum.

\begin{figure*}
  \vskip -.4 in
  \hskip .3 in
  \resizebox{7.5in}{!}{\includegraphics{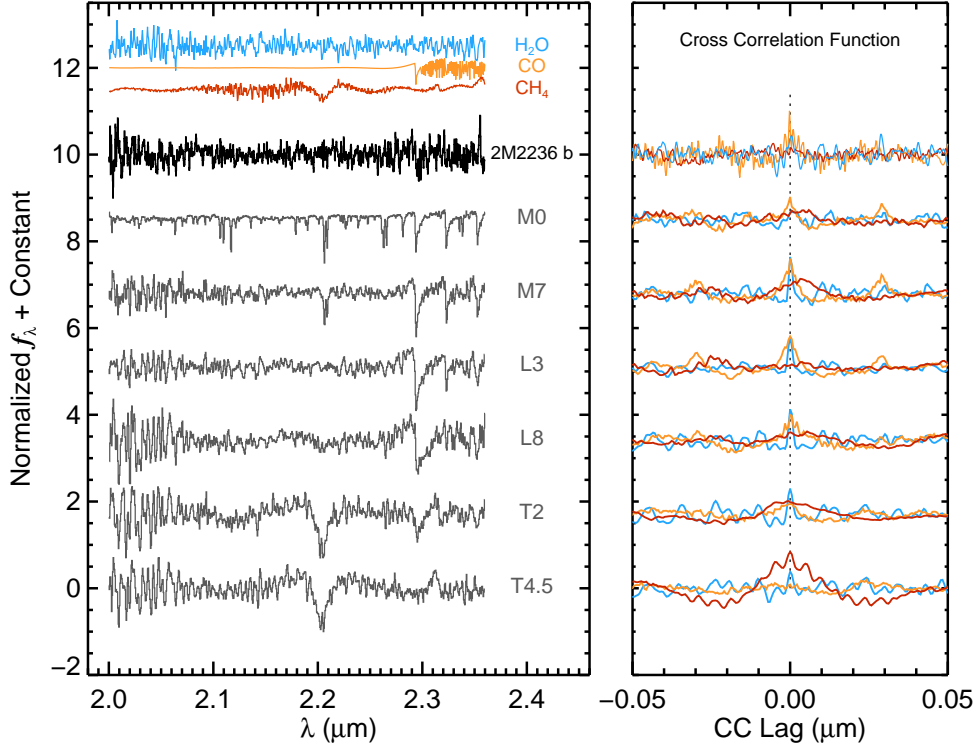}}
  \vskip -1 in
  \caption{Cross correlation of pure H$_2$O, CO, and CH$_4$ templates with the flattened $K$-band spectrum of 
  2M2236+4751\,b and MLT templates from the IRTF Spectral Library.  The cross correlation function (CCF) 
  power for CO (orange) 
  is strong in M and L dwarfs then diminishes for T dwarfs as methane (red) becomes the dominant 
  carbon carrier.  Water absorption (blue) is evident in all objects but is noticeably weaker in the 
  mid-T dwarf spectrum.  For 2M2236+4751\,b, CO is prominent in the CCF, water is weak, and 
  we find no evidence for methane despite its cool effective temperature of $\approx$900--1200~K. \label{fig:molxcor} } 
\end{figure*}

Results of the cross correlation are shown in Figure~\ref{fig:molxcor} along with 
the same analysis for M, L, and T dwarfs from the IRTF Spectral Library 
(\citealt{Cushing:2005ed}; \citealt{Rayner:2009ki}).  
Water shows a significant peak with no cross correlation lag (velocity shift) in all MLT templates.
As expected, the transition from L to T dwarfs is marked by a reduced strength in 
CO power and an increase in broad methane cross correlation peak.
Although the M dwarfs show substantial power from the CH$_4$ template, this is a result of 
the strong 2.2~$\mu$m methane absorption feature locking onto the 2.21 $\mu$m \ion{Na}{1}
doublet at high temperatures.
For 2M2236+4751\,b, we find strong evidence of CO, some evidence of H$_2$O, and no
sign of methane absorption.  The slight trough from 2.2--2.3~$\mu$m in our OSIRIS spectrum
is therefore probably not a result of methane absorption.


We estimate a bolometric luminosity for 2M2236+4751\,b using the 0.9--2.4~$\mu$m spectrum of 
PSO~J318.5--22 from \citet{Liu:2013gya}--- which bears a close resemblance to 
2M2236+4751\,b--- flux-calibrated to the $K_S$-band
photometry of 2M2236+4751\,b together with a [$T_\mathrm{eff}$=1100~K; $\log g$=4.5~dex] 
BT-Settl atmospheric model
from \citet{Baraffe:2015fwa} as a bolometric correction at short ($\lambda$ $<$0.9~$\mu$m) 
and long ($\lambda$ $>$ 2.4~$\mu$m) wavelengths.
\citet{Filippazzo:2015dv} show that the flux redistribution from short to long wavelengths 
for young L dwarfs pivots between
1.5--2.5 $\mu$m, so the atmospheric model parameters for our luminosity calculation 
are chosen to correspond 
to the $M_H$- and $M_{K_S}$-band magnitudes of 2M2236+4751\,b predicted by Cond
evolutionary models (\citealt{Baraffe:2003bj}).\footnote{Dusty evolutionary models (\citealt{Chabrier:2000hq})
are truncated at 0.012~\Msun \ for an age of 120~Myr but imply similar upper limits on the physical properties of 2M2236+4751\,b: 
$T_\mathrm{eff}$$<$1300~K, $\log g$ $<$ 4.2~dex, and $M$$<$12.5~\Mjup.}
This effective temperature is consistent with the value from the \citet{Filippazzo:2015dv} 
young $T_\mathrm{eff}$($M_H$) relation, which yields 
970 $\pm$ 150~K for 2M2236+4751\,b.
Uncertainties in $L_\mathrm{bol}$ are derived in a Monte Carlo 
fashion by randomly and repeatedly sampling new distances and absolute
flux calibration scale factors from Normal distributions based on our kinematic distance estimate
(63 $\pm$ 5 pc) and apparent $K_S$ magnitude
of 2M2236+4751\,b (13.35 $\pm$ 0.18 mag).
We estimate a luminosity of log($L_\mathrm{bol}$/$L_{\odot}$) = --4.57 $\pm$ 0.06 dex from 
the mean and standard deviation of 10$^4$ trials.

The inferred mass of 2M2236+4751\,b is 13.3~$\pm$~0.8 \Mjup \ based on its age, its 
bolometric luminosity, and the ``hybrid'' evolutionary models of \citet{Saumon:2008im}.
(The corresponding effective temperature from these models is 1170~$\pm$ 40~K.)
Using absolute magnitudes instead of bolometric luminosity, we find masses of $\approx$8~\Mjup,
$\approx$11~\Mjup, and $\approx$11~\Mjup \  for $M_J$, $M_H$, and $M_{K_S}$-band magnitudes based on the
Cond evolutionary models of \citet{Baraffe:2003bj}.  
Dusty models from \citet{Chabrier:2000hq} imply an upper limit of $<$12.5~\Mjup.
Altogether we adopt a mass range 
of 11--14~\Mjup \ for 2M2236+4751\,b assuming the system is a member of AB Dor.  
In the unlikely case  it is a kinematic interloper, the inferred mass of the companion is 43 $\pm$ 5~\Mjup,
70 $\pm$ 2~\Mjup, and 74 $\pm$ 1~\Mjup \ at ages of 1 Gyr, 5 Gyr, and 10 Gyr.  The corresponding effective temperatures are 1310 $\pm$ 50 K, 
1390 $\pm$ 40 K, and  1390 $\pm$ 40 K for the same ages.
Additional photometry--- especially at mid-infrared wavelengths---
will help to better constrain the luminosity and mass of this remarkable companion.

\section{Discussion}{\label{sec:discussion}}

Empirical sequences of substellar isochrones in color-magnitude diagrams 
are important tools to chart the physical and spectral properties of brown dwarfs and giant planets at a given age,
 map their entire cooling history at a given mass, 
and jointly test low-temperature atmospheric and evolutionary models.
The positions of young ($\sim$10--200~Myr) substellar objects
in near-infrared color-magnitude diagrams 
have progressively come into focus
with programs searching for  
low-mass members of young moving groups (\citealt{Gagne:2014gp}; \citealt{Aller:2016kg}) 
and gas giant planets with high-contrast imaging (\citealt{Bowler:2016jk})
together with follow-up parallax programs to measure distances 
(\citealt{Faherty:2016fx}; Liu et al. 2016).  
This is particularly true for the young late-M through mid-L dwarf populations, 
whose numbers and trigonometric distances have swelled in recent years,
but the location of young late-L, L/T, and T dwarfs have largely remained elusive owing to the relative 
dearth of discoveries in this
regime and their intrinsic scarcity.
 
\begin{figure*}
  \vskip -1.2 in
  \hskip -.2 in
  \resizebox{7.5in}{!}{\includegraphics{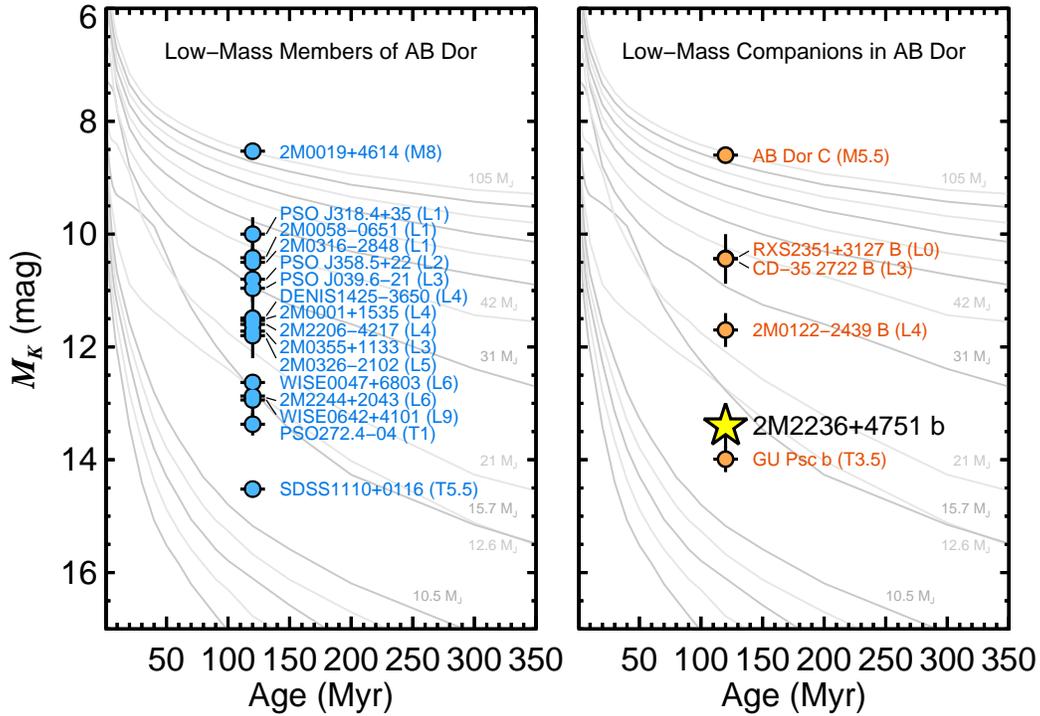}}
  \vskip -.6 in
  \caption{Confirmed and candidate ultracool members of the AB Dor young moving group.  In absolute $K$-band magnitude,
  2M2236+4751\,b (yellow star) is fainter than all isolated brown dwarfs (left) except the T5.5 object 
  SDSS J1110+0116 (\citealt{Gagne:2015kf}), and among companions it is the second lowest-mass
  member after GU~Psc~b (\citealt{Naud:2014jx}).  BT-Settl evolutionary models are 
  depicted in gray (\citealt{Baraffe:2015fwa}).    \label{fig:abdorcomp} } 
\end{figure*}

The AB Dor moving group is among the oldest of the 
nearby ($\lesssim$100~pc) young ($\lesssim$200~Myr) comoving associations 
in the solar neighborhood (e.g., \citealt{Zuckerman:2004ku}; 
\citealt{Torres:2008vq}).  With an age of $\sim$120~$\pm$~10~Myr, 
it straddles younger well-characterized moving groups like $\beta$~Pic and Tuc-Hor at 20--50~Myr
and older, more heavily populated clusters at 500--800~Myr like 
Ursa Majoris, the Hydes, and Coma Ber (\citealt{Mamajek:2016ik}), thereby
serving as an important benchmark for stellar and substellar evolution.
The population of confirmed or suspected free-floating brown dwarfs in AB Dor now includes over a dozen
objects spanning the entire 
L dwarf sequence down to $\sim$T1 (e.g., \citealt{Best:2015em}; \citealt{Aller:2016kg}) 
as well as the single mid-T dwarf candidate SDSS J1110+0116 (\citealt{Gagne:2015kf}).
Many brown dwarf and planetary-mass companions have also been identified 
with masses ranging from roughly 9 to 40~\Mjup \ (Figure~\ref{fig:abdorcomp}):
1RXS~J2351+3127 B (L0; \citealt{Bowler:2012cs}), CD--35 2722 B (L3; \citealt{Wahhaj:2011by}),
2M0122--2439~B (L4; \citealt{Bowler:2013ek}), and GU~Psc b (T3.5; \citealt{Naud:2014jx}).
Altogether, this is the largest set of substellar companions for any young moving group.
 
\begin{figure}
  \vskip .1 in
  \hskip -.1 in
  \resizebox{3.3in}{!}{\includegraphics{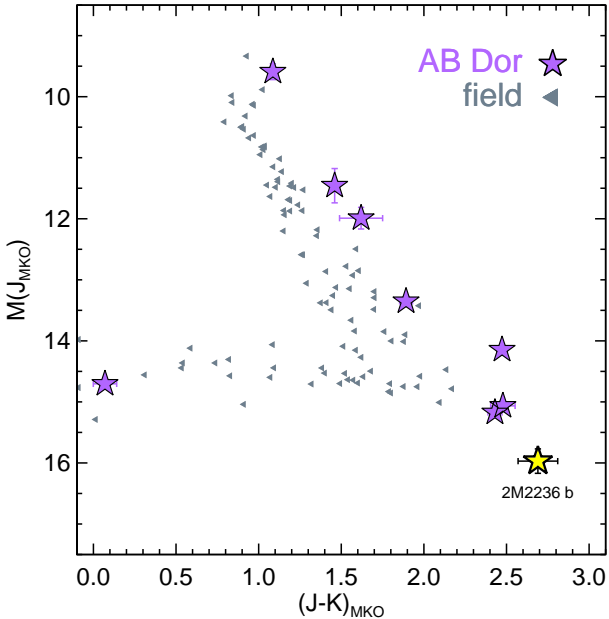}}
  \vskip 0 in
  \caption{Color-magnitude diagram showing the position of 2M2236+4751\,b compared to ultracool members of the AB Dor moving group with parallactic distances and high membership probability as summarized by Liu et al. (2016).  We also include PSO~J318.4243+35.1277 from \citet{Aller:2016kg}, though we do not plot the other two objects in that paper with parallaxes (PSO~J039.6352--21.7746 and PSO~J358.5527+22.1393) as radial velocity followup by Aller (2016, PhD thesis) leads to inconclusive membership results.  
Normal ultracool field objects are shown as small gray triangles (from \citet{Dupuy:2012bp} for objects with apparent magnitude errors $<$0.10 mag in $J$ and $K$ and with $J$-band absolute magnitude errors of $<$0.10~mag.  
   \label{fig:abdoriso} } 
\end{figure}

With a mass of 11--14~\Mjup, 2M2236+4751\,b is the second lowest-mass companion discovered in 
AB Dor after GU Psc~b (11 $\pm$ 2~\Mjup; Figure~\ref{fig:abdorcomp}).  2M2236+4751\,b
is also the reddest known member of this group and resides in a prominent position  
at the ``elbow'' of its substellar isochrone (Figure~\ref{fig:abdoriso}).
It has an even more extreme color than two 
red free-floating brown dwarfs in AB Dor, WISE~J0047+6803 (($J$--$K$)$_\mathrm{MKO}$= 2.48 $\pm$ 0.08 mag) 
and 2M2244+2043 (($J$--$K$)$_\mathrm{MKO}$ = 2.43 $\pm$ 0.04 mag),  
demonstrating that the tip of the L dwarf 
sequence extends to at least ($J$--$K$)$_\mathrm{MKO}$ = 2.7~mag 
even at the relatively old age of this cluster.


\begin{figure*}
  \vskip -0.1 in
  \hskip .6 in
  \resizebox{5.5in}{!}{\includegraphics{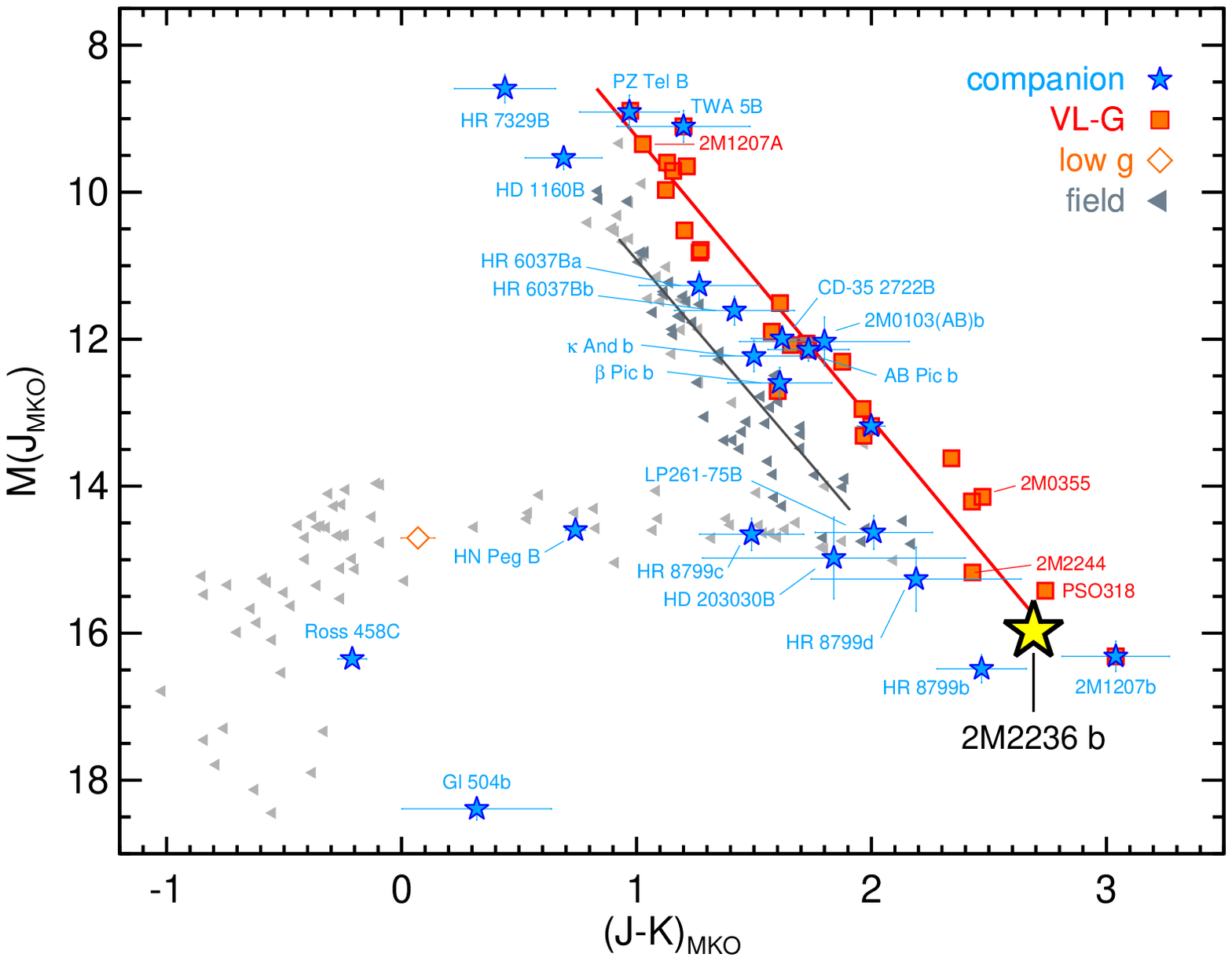}}
  \vskip 0 in
  \caption{The ultracool $J_\mathrm{(MKO)}$/($J$--$K$)$_\mathrm{(MKO)}$ color-magnitude diagram (adapted from Liu et al. 2016).
   Triangles denote field brown dwarfs from \citet{Dupuy:2012bp}, filled squares show late-M and L
   dwarfs with very-low gravity classifications (Liu et al. 2016), and blue stars represent companions with trigonometric
   distances.  2M2236+4751\,b (yellow star) is located at the tip of the faint red L dwarf sequence in a sparsely populated
   region characterized by extreme cloud properties.  Gray and red lines show the linear fit of \fldg \ and \vlg \ 
   brown dwarfs from Liu et al. (2016).    \label{fig:cmd} } 
\end{figure*}

The shape of the AB Dor substellar isochrone has important implications for other clusters with similar ages. 
The Pleiades is an especially interesting example with a comparable age of about 125~Myr (\citealt{Stauffer:1998kt}) 
and possible dynamical relationship with the AB Dor group (\citealt{Ortega:2007go}).
Although it is one of the best-characterized nearby open clusters with a population of over one 
thousand stars (e.g., \citealt{Stauffer:2007gf}), 
its very low-mass substellar members have been difficult to explore owing to their intrinsic faintness.
Nevertheless, several deep near-infrared imaging surveys have identified L dwarfs down to the planetary-mass regime (e.g., \citealt{Bihain:2006ck}; \citealt{Lodieu:2007gd}).
\citet{ZapateroOsorio:2014em} show the Pleiades L dwarf sequence in the $J$/$J$--$K$ 
diagram generally resembles AB Dor, progressively 
reddening to $J$--$K$ values of $\sim$2.5~mag, but then appears to turn over to bluer colors beyond $J$$\sim$20.3~mag ($M_J$$\sim$14.7~mag) 
and $K_S$$\sim$17.8~mag ($M_{K_S}$$\sim$12.2~mag; \citealt{ZapateroOsorio:2014dr}).
However, the absolute magnitudes of 2M2236+4751\,b ($M_J$=15.97 $\pm$ 0.2~mag; $M_{K_S}$=13.35 $\pm$ 0.18~mag) 
are over one magnitude \emph{fainter} than the location of this apparent turnover.
Similarly, the two aforementioned red L dwarfs WISE~J0047+6803 and 2M2244+2043 in AB Dor 
have absolute magnitudes nearly half a magnitude fainter than this possible L/T turnover, 
suggesting that there is either an age discrepancy between AB Dor and the Pleiades, 
a dramatic level of photospheric diversity exists among brown dwarfs with similar absolute magnitudes at this age,
or that some of the faint blue planetary-mass candidates identified by \citet{ZapateroOsorio:2014em} are contaminants.

Compared to other companions, 2M2236+4751\,b appears to have even more extreme photospheric
properties than all previously known objects except the $\approx$5~\Mjup \ object 
2M1207--3932~b (($J$--$K$)$_\mathrm{MKO}$ = 3.0 $\pm$ 0.2 mag; Figure \ref{fig:cmd}).
However, all isolated objects and companions in this region of the color-magnitude diagram
show qualitatively similar atmospheric properties: red colors, faint absolute magnitudes, 
strong CO absorption, and no signs of methane absorption despite having effective temperatures 
below the traditional L/T transition of $\approx$1200--1400~K for the older field population (e.g., \citealt{Golimowski:2004en}).
\citet{Barman:2011dq} and \citet{Skemer:2014hy} find that very thick clouds and 
rapid vertical mixing are required to reproduce the red colors and lack of methane 
absorption in 2M1207--3932~b, the most extreme object at the tip of the L dwarf sequence,
and these traits are likely to apply to 2M2236+4751\,b as well.
However, Liu et al. (2016) show that another population of red L dwarfs like 2M2148+4003 (\citealt{Looper:2008hs})
and WISE~J2335+4511 (\citealt{Thompson:2013kv})
which do not appear to be young can also reside in this part of the color-magnitude diagram,
indicating that surface gravity is not the sole underlying explanation for 
the shared atmospheric traits of these objects.
2M2236+4751\,b reinforces the impression that there do not appear to be any obvious differences between the red, 
faint objects being discovered as companions and the analogous population of isolated free-floating objects.

With a projected separation of 230~AU, 2M2236+4751\,b is the latest example of a growing population of 
planetary-mass companions on wide orbits of several hundred AU (e.g., \citealt{Bailey:2014et}; \citealt{Kraus:2014tl}; \citealt{Deacon:2016dg}).
The origin of these objects is unclear; planet-planet scattering (\citealt{Veras:2009br}), the tail end of disk instability (\citealt{Kratter:2010gf}), and
turbulent fragmentation of molecular clouds (\citealt{Bate:2009br}) 
have all been invoked to explain their origin.  
\citet{Bryan:2016vg} argue against scattering as the dominant origin of this population due to the 
lack of additional close-in companions in these systems and the dearth of  multiple massive ($\sim$10~\Mjup) planets uncovered
at small separations in radial velocity surveys.
In fact, because of their low numbers and lack of robust statistics,
it is unclear whether these objects make up their own population or are part of a broader distribution of
giant planets or low-mass brown dwarfs spanning small separations (tens of AU) out to extremely wide separations (thousands of AU).
Unlike most of the wide planetary-mass companions currently known,
2M2236+4751\,b is part of a well-defined survey 
(to be discussed in more detail in a future publication) and will help clarify the statistical properties of 
planetary-mass companions on wide orbits around low-mass stars.

\section{Summary and Conclusions}{\label{sec:conclusions}}

We have presented the discovery of a faint red L dwarf companion to the late K dwarf 2MASS~J22362452+4751425.
The radial velocity and proper motion of the host star are consistent with the 120~Myr AB Dor young moving group.
2M2236+4751\,A is relatively inactive, showing partly-filled H$\alpha$ absorption but no X-ray emission or UV excess,
suggesting that it is either in the tail end of the AB Dor activity distribution or it is an older kinematic interloper in that group.
The unusually red colors of the companion are similar to (but more extreme than) other mid- to late-L dwarfs in AB Dor,
possibly bolstering membership likelihood, but the existence of high-gravity brown dwarfs in this part of the color-magnitude diagram
means that this trait is not uniquely a sign of younth and low surface gravity.
Assuming the pair is indeed a member of AB Dor, the mass of 
2M2236+4751\,b is 11--14~\Mjup \ from hot start evolutionary models
and resides at a projected separation of $\approx$230~AU at the kinematic distance of $\approx$65~pc to the host star.
If the system is a older kinematic interloper then the implied mass of the companion can be as high as $\sim$74~\Mjup \ at 10~Gyr.

The near-infrared colors of 2M2236+4751\,b are among the reddest known of any brown dwarf or giant planet.
In near-infrared color-magnitude diagrams, it appears to mark the elbow of the AB Dor substellar isochrone between
other faint red L dwarfs like WISE~J0047+6803 and 2M2244+2043 and the two T dwarfs GU~Psc~b and SDSS J1110+0116.
This implies that the transition from red L dwarfs to dust-free T dwarfs occurs between 
a narrow mass range of about 11--13~\Mjup \ at an age of $\approx$120~Myr.
The colors of 2M2236+4751\,b are redder than HR~8799~b but slightly bluer than 2M1207--3932~b, demonstrating that 
even intermediate-age giant planets can possess similarly extreme atmospheric properties beyond 100~Myr.

The 2M2236+4751\,Ab system is well-suited for a broad range of follow-up studies.
The age and mass of 2M2236+4751\,b will eventually be clarified with a parallax measurement of the host star from $Gaia$ to 
establish whether the system is a member of AB Dor.  
Follow-up photometry of the companion at both shorter and longer wavelengths will better constrain its luminosity, spectral type, and 
atmospheric properties.
Both low- and high-resolution near-infrared spectroscopy of 2M2236+4751\,b can be used to derive spectral and gravity
classifications, independently age-date the system through gravity-dependent absorption line depths, measure
its projected rotational velocity, and determine its composition and abundance ratios.
Finally, 2M2236+4751\,b is also an excellent target for variability studies to measure its rotational period and search for signatures of 
heterogeneity and patchiness in its unusually thick clouds.  

\acknowledgements

We thank A. Kraus for helpful discussions about the age of the host star and the 
entire Keck Observatory staff for their exceptional support. 
I.T. was supported by the Department of Astronomy at the University of Texas at Austin
though the Cox Endowment and Board of Visitors Funds,
as well as a NASA WIYN PI Data Award administered by the NASA Exoplanet Science Institute 
as part of NN-EXPLORE through the scientific 
partnership of the National Aeronautics and Space Administration, 
the National Science Foundation, and the National Optical Astronomy Observatory. 
This paper includes data taken at The McDonald Observatory of The University of Texas at Austin.
Based on observations obtained with ESPaDOnS, located at the Canada-France-Hawaii Telescope (CFHT). CFHT is operated by the National Research Council of Canada, the Institut National des Sciences de l'Univers of the Centre National de la Recherche Scientique of France, and the University of Hawai'i. ESPaDOnS is a collaborative project funded by France (CNRS, MENESR, OMP, LATT), Canada (NSERC), CFHT and ESA. We utilized data products from the Two Micron All Sky Survey, which is a joint project of the University of Massachusetts 
and the Infrared Processing and Analysis Center/California Institute of Technology, funded by the National Aeronautics and 
Space Administration and the National Science Foundation.  NASA's Astrophysics Data System Bibliographic Services together 
with the VizieR catalogue access tool and SIMBAD database operated at CDS, Strasbourg, France, were invaluable resources for this work.  
This work used the Immersion Grating Infrared Spectrometer (IGRINS) that was developed under a collaboration between the 
University of Texas at Austin and the Korea Astronomy and Space Science Institute (KASI) with the financial support of the 
US National Science Foundation under grant AST- 1229522, to the University of Texas at Austin, and of the Korean GMT Project of KASI.
This publication makes use of data products from the Wide-field Infrared Survey Explorer, which is a joint project of the University of California, 
Los Angeles, and the Jet Propulsion Laboratory/California Institute of Technology, funded by the National Aeronautics and Space Administration.
Finally, mahalo nui loa to the kama`\={a}ina of Hawai`i for their support of Keck and the Mauna Kea observatories.  
We are grateful to conduct observations from this mountain.

\facility{{\it Facilities}: \facility{Keck:II (NIRC2)}, \facility{Keck:I (OSIRIS)}, \facility{Smith (IGRINS)}, \facility{IRTF (SpeX)}, \facility{CFHT (ESPaDOnS)}}

\bibliographystyle{apj}

\end{document}